
\input phyzzx

\overfullrule=0pt
{}~\hfill\vbox{\hbox{TIFR-TH-92/03}\hbox{January, 1992}}

\title{ON THE BACKGROUND INDEPENDENCE OF STRING FIELD THEORY}
\titlestyle{III. Explicit Field Redefinitions}

\author{Ashoke Sen}

\address{Tata Institute of Fundamental Research, Homi Bhabha Road, Bombay
400005, India}

\abstract

Given two conformal field theories related to each other by a marginal
perturbation, and string field theories constructed around such
backgrounds, we show how to construct explicit redefinition of string
fields which relate these two string field theories.
The analysis is carried out completely for quadratic and cubic terms in
the action.
Although a general proof of existence of field redefinitions which relate
higher point vertices is not given, specific examples are discussed.
Equivalence of string field theories formulated around two conformal field
theories which are not close to each
other, but are related to each other by a series of marginal deformations,
is also discussed.
The analysis can also be applied to study the equivalence of different
formulation of string field theories around the same background.

\endpage

\def\r{\rangle}   \def\bL{\bar L} \def\btL{\bar{\tilde L}}
\def\l{\langle}    \def\p{\partial}
\def\psr{|\Psi\r}
\def\tpsr{|\tilde\Psi\r}    \def\tPs{\tilde\Psi} \def\tPh{\tilde\Phi}
\def\tS{\tilde S} \def\hS{\hat S} \def\hP{\hat\Psi} \def\hp{\hat\psi}
\def\tps{\tilde\psi} \def\hA{\hat A}
\def\ch{{\cal H}}
\def\tch{\tilde{\cal H}}  \def\tL{\tilde L}
\def\tqb{\tilde Q_B}  \def\rpp{\r''}
\def\tA{\tilde A} \def\pcl{\Psi_{cl}}
\def\TWO{{[2]}}
\def\zero{{(0)}}  \def\one{{(1)}} \def\two{{(2)}} \def\en{{(N)}}
\def\no{{(N-1)}} \def\em{{(M)}} \def\mo{{(M-1)}} 
 \def\thr{{(3)}} \def\four{{(4)}} \def\five{{(5)}}
\def\vp{\varphi} \def\cs{{\cal S}} \def\hqb{\hat Q_B}
\def\bp{\bar\psi} \def\bS{\bar S} \def\bA{\bar A} \def\bP{\bar\Psi}
\def\at{{\alpha_2}} \def\bt{{\beta_2}} \def\gt{{\gamma_2}}
\def\dt{{\delta_2}} \def\mt{\mu_2} \def\nt{{\nu_2}} \def\tt{{\tau_2}}
\def\st{{\sigma_2}} \def\ao{{\alpha_1}} \def\tl{\tilde\lambda}
\def\tB{\tilde B} \def\bB{\bar B} \def\bl{\bar\lambda} \def\tV{\tilde V}
\def\kt{{k_2}} \def\tG{\tilde\Gamma} \def\bG{\bar\Gamma} \def\lt{l_2}
\def\co{{\cal O}} \def\bc{\bar c} \def\tV{\tilde V} \def\sk{|s(k)\r}
\def\ba{\bar\alpha}

\let\refmark=\NPrefmark 

\def\define#1#2\par{\def#1{\Ref#1{#2}\edef#1{\noexpand\refmark{#1}}}}
\def\con#1#2\noc{\let\?=\Ref\let\<=\refmark\let\Ref=\REFS
         \let\refmark=\undefined#1\let\Ref=\REFSCON#2
         \let\Ref=\?\let\refmark=\<\refsend}

\define\SCHUBERT
C. Schubert, MIT preprint CTP 1977.

\define\LPP
A. LeClair, M. Peskin and C. Preitschopf, Nucl. Phys. {\bf B317} (1989)
411,464.

\define\CUBIC
G. T. Horowitz, J. Lykken, R. Rohm and A. Strominger, Phys. Rev. Lett.
{\bf 57} (1986) 283.

\define\BATALIN
I.~A.~Batalin and G.~A.~Vilkovisky, Phys.~Lett. {\bf 102B} (1981) 27; {\bf
120B} (1983) 166; Phys.~Rev. {\bf D28} (1983) 2567; J.~Math.~Phys. {\bf
26} (1985) 172.

\define\RENORM
A. Sen, Phys.~Lett. {\bf B252} (1990) 566.

\define\SIEGELBOOK
W. Siegel, Introduction to String Field Theory, World Scientific, 1988.

\define\NELSON
M. Campbell, P. Nelson and E. Wong, Int. J. Mod. Phys. {\bf A6} (1991)
4909.

\define\THORN
C. Thorn,  Nucl.~Phys. {\bf B286} (1987) 61.

\define\GHMU
D. Ghoshal and S. Mukherji, Mod. Phys. Lett. {\bf A6} (1991) 939.

\define\BPZ
A.~Belavin, A.~M.~Polyakov and A.~B.~Zamolodchikov, Nucl.~Phys. {\bf B241}
(1984) 333.

\define\GAUGEGRAV
D. Ghoshal and A. Sen, preprint TIFR-TH-91/47.

\define\SENPOLY
A.~Sen, Phys.~Lett. {\bf B241} (1990) 350.

\define\KAKU
M. Kaku and J. Lykken, Phys. Rev. {\bf D38} (1988) 3067;
M. Kaku, preprints CCNY-HEP-89-6, Osaka-OU-HET 121.

\define\NONPOL
M.~Saadi and B.~Zwiebach, Ann.~Phys. {\bf 192} (1989) 213;
T.~Kugo, H.~Kunitomo, and K.~Suehiro, Phys.~Lett. {\bf 226B} (1989) 48;

\define\GAUGEINV
T.~Kugo and K.~Suehiro, Nucl.~Phys. {\bf B337} (1990) 434.

\define\SYCL
S.~Mukherji and A.~Sen,  Nucl. Phys. {\bf B363} (1991) 639.

\define\SOZW
H. Sonoda and B. Zwiebach, Nucl. Phys. {\bf B331} (1990) 592.

\define\HLOOP
H.~Hata, Phys.~Lett. {\bf 217B} (1989) 438, 445; Nucl.~Phys. {\bf B329}
(1990) 698; {\bf B339} (1990) 663.

\define\SAADI
M. Saadi, Mod. Phys. Lett. {\bf A5} (551) 1990; Int. J. Mod. Phys. {\bf
A6} (1991) 1003.

\define\ZWIEBACH
B.~Zwiebach, Mod.~Phys.~Lett. {\bf A5} (1990) 2753;
Phys. Lett. {\bf B241} (1990) 343; Comm. Math. Phys. {\bf 136} (1991) 83.

\define\MKAKU
L. Hua and M. Kaku, Phys. Lett. {\bf B250} (1990) 56;
M. Kaku, Phys. Lett. {\bf B250} (1990) 64.

\define\SENFIELD
A.~Sen, Nucl.~Phys. {\bf B334} (1990) 350, 395; {\bf B335} (1990) 435.

\define\BACKGND
A.~Sen, Nucl.~Phys. {\bf B345} (1990) 551.

\define\BACKTWO
A.~Sen, Nucl.~Phys. {\bf B347} (1990) 270.

\chapter{INTRODUCTION}

A complete formulation of closed bosonic string field theory has been
given in the last few years in terms of non-polynomial
interactions\NONPOL\GAUGEINV\ (see also ref.\KAKU), and quantization of
this
string field theory has also been carried
out\con\HLOOP\SOZW\SAADI\ZWIEBACH\MKAKU\noc.
Such a field theory can be formulated not only in the background of flat
space-time, but also in the background of any arbitrary conformal field
theory\SENPOLY\ following the formulation of refs.\LPP\SENFIELD.
The natural question to ask is whether the string field
theory is background independent, i.e. given the string field theory
action around two different conformal field theory backgrounds, whether we
can find a redefinition of string fields which relates these two actions.

The question was partially answered in two previous
papers\BACKGND\BACKTWO.
In these papers we studied the relationship between string field theories
formulated around two neighbouring conformal field theories, $-$ CFT and
CFT$''$, $-$ related by a marginal perturbation.
Let $\Psi$ and $\tPs$ denote the string fields corresponding to string
field theories formulated around CFT and CFT$''$ respectively, and
$S(\Psi)$ and $\tS(\tPs)$ be the corresponding string field theory
actions.
It was shown in ref.\BACKGND\ that  there is a classical solution $\pcl$
of the equations of motion $(\p S/\p \Psi) =0$ such that if we define $\hP
= \Psi -\pcl$ and $\hS(\hP) = S(\Psi) - S(\pcl)$, then the kinetic
operator of $\hS(\hP)$ is related to the kinetic operator of $\tS(\tPs)$
by a similarity transformation.
Furthermore, the linearized form of the gauge transformation of $\hP$ is
also related to that of $\tPs$ by the same similarity transformation.
The analysis was carried out to the first order in the perturbation which
relates CFT and CFT$''$.
In other words, ref.\BACKGND\ established the equivalence between the
quadratic terms in the actions $\hS(\hP)$ and $\tS(\tPs)$.
On the other hand, in ref.\BACKTWO\ we analyzed the physical $S$-matrix
elements in the two theories described by the actions $\hS(\hP)$ and
$\tS(\tPs)$, and showed that they are the same.
The analysis was carried out for all three point amplitudes and all
$N$-point tachyonic amplitudes.

In this paper, we shall show with the help of the results of
refs.\BACKGND\
and \BACKTWO\ that one can construct explicit redefinition of
string fields which converts the action $\tS(\tPs)$ to $\hS(\hP)$, and
hence to $S(\Psi)$.
The analysis is carried out completely for the  cubic vertices in the two
theories, using a method similar to the one used in ref.\GAUGEGRAV.
(The analysis for the quadratic terms was already done in ref.\BACKGND).
We also discuss some specific features that appear in the analysis of the
higher order vertices, and show, in special cases, how explicit field
redefinitions may be found which relate the vertices in $\tS(\tPs)$ and
$\hS(\hP)$.
But a general proof of equivalence of these two vertices is not given.

Assuming  that a complete set of field redefinitions can be found which
relate $S(\Psi)$ and $\tS(\tPs)$ to first order in the perturbation
parameter $\lambda$, one can ask if the analysis can be extended beyond
first order in $\lambda$ and hence can be used to relate string field
theories around backgrounds that are not necessarily close to each other
but can be
obtained from each other by a series of marginal deformations.
Intuitively it is clear that the string field theories constructed around
two such backgrounds will also be related by field redefinition, which
can be built by combining successive field redefinitions which relate two
nearby conformal field theories.
We give a general algorithm for finding these finite field redefinitions
in terms of the infinitesimal ones.

The plan of the paper is as follows.
Sect.2 contains a precise formulation of the problem that we are going to
study, as well as its relationship with the work of ref.\BACKGND.
In sect.3 we prove the existence of field redefinitions that relate the
cubic vertices of these two theories, and also give the general algorithm
for constructing these field redefinitions.
Sect.4 contains a discussion of the corresponding analysis for higher
point vertices.
Sect.5 deals with the case where the two conformal field theories are not
necessarily close.
We also use the result of this section to construct the classical solution
in string field theory representing the perturbed conformal field theory
to second order in the perturbation parameter.
We conclude in sect.6 with a discussion of our results and some comments.
The three appendices contain some of the technical results needed in the
analysis of sects. 3, 4 and 5.

\chapter{FORMULATION OF THE PROBLEM}

We begin this section with a precise formulation of the problem that we
want to solve.
Let CFT and CFT$''$ be two different conformal field theories, both with
central charge 26, and hence both providing consistent background for the
formulation of string theory.
Let $\psr$ and $\tpsr$ denote string fields for the string field theories
formulated around CFT and CFT$''$ respectively, and $S(\Psi)$ and
$\tS(\tPs)$ be the actions of the corresponding string field theories.
In order to prove background independence of string field theory, one
needs to find a functional relationship of the form $\Psi = f(\tPs)$ such
that,
$$
S(f(\tPs)) =\tS(\tPs) + {\rm constant}
\eqn\etwofive
$$
Let us denote by CFTG (CFTG$''$) the combined conformal field theory of the
ghost system and CFT (CFT$''$).
If $\ch$ ($\tch$) denote the complete Hilbert space of CFTG (CFTG$''$), and
$L_n$, $\bL_n$ ($\tL_n$, $\bar{\tL_n}$) denote the total Virasoro
generators of CFTG (CFTG$''$),
then $b_0^-\psr$ ($b_0^-\tpsr$) is an arbitrary state in $\ch$ ($\tch$) with
ghost number 2 and annihilated by $b_0^-$ and $L_0^- \equiv L_0-\bL_0$
($\tL_0 \equiv \tL_0 - \btL_0$).
Then the actions $S(\Psi)$ and $\tS(\tPs)$ are given by,
$$
S(\Psi) = {1\over 2} \l\Psi | Q_B b_0^- |\Psi\r + \sum_{N=3}^\infty
{g^{N-2}\over N!} \{\Psi^N\}
\eqn\etwoone
$$
and
$$
\tS(\tPs) ={1\over 2} \l\tPs |\tqb b_0^- |\tPs\rpp +\sum_{N=3}^\infty
{g^{N-2}\over N!} \{\tPs^N\}''
\eqn\etwotwo
$$
with the gauge invariance,
$$
b_0^- \delta \psr = Q_B b_0^- |\Lambda\r +\sum_{N=3}^\infty {g^{N-2}\over
(N-2)!} [\Psi^{N-2}\Lambda]
\eqn\etwothree
$$
$$
b_0^- \delta\tpsr = \tqb b_0^- |\tilde\Lambda\r +\sum_{N=3}^\infty
{g^{N-2}\over (N-2)!} [\tPs^{N-2}\tilde\Lambda]''
\eqn\etwofour
$$
where $b_0^-|\Lambda\r$ ($b_0^-|\tilde\Lambda\r$) are states of ghost
number 1 in $\ch$ ($\tch$) annihilated by $b_0^-$ and $L_0^-$ ($\tL_0^-$).
$Q_B$ ($\tqb$) is the nilpotent BRST operator acting in $\ch$ ($\tch$),
and $\l|\r$ ($\l|\rpp$) denote the BPZ inner product\BPZ\ in CFTG
(CFTG$''$).
Thus the operators $L_n$, $Q_B$ ($\tL_n$, $\tqb$) etc. have appropriate
hermiticity properties with respect to the inner product $\l|\r$
($\l|\r''$) but not with respect to the inner product $\l|\r''$ ($\l
|\r$).
$[A_1\ldots A_N]$ and
$\{A_1\ldots A_N\}\equiv (-1)^{n_1+1}\l A_1|[A_2\ldots A_N]\r$
denote multilinear maps from
$N$-fold tensor product of $\ch$ to  $\ch$ and $C$ respectively, and are
constructed in terms of correlation functions in
CFTG\LPP\NONPOL\GAUGEINV\SENPOLY\BACKGND\BACKTWO.
Here $n_1$ denotes the ghost number of the state $|A_1\r$.
Similarly,  $[\tA_1\ldots \tA_N]''$ and
$\{\tA_1\ldots \tA_N\}''\equiv (-1)^{n_1+1} \l \tA_1|
[\tA_2\ldots \tA_N]''\r''$
denote
multilinear maps from $N$-fold tensor product of $\tch$ to  $\tch$ and $C$
respectively, and are constructed in terms of correlation functions in
CFTG$''$.

Let us introduce a basis of states $|\Phi_{n,r}\r$ in the subspace $\ch_n$
of $\ch$ of ghost number $n$, and annihilated by $b_0^-$ and $L_0^-$.
Similarly, let $|\tPh_{n,r}\r$ be a basis of states in the subspace
$\tch_n$ of $\tch$ of ghost number $n$ and annihilated by $b_0^-$ and
$\tL_0^-$.
Then we may write,
$$
b_0^-|\Psi\r = \sum_r \psi_r |\Phi_{2,r}\r
\eqn\etwosix
$$
$$
b_0^-\tpsr = \sum_r\tps_r|\tPh_{2,r}\r
\eqn\etwoseven
$$
It was shown in ref.\BACKGND\ that the operators $\tL_m$, $\bar{\tL}_m$ in
$\tch$ may be
represented in the Hilbert space $\ch$ and vice versa, that is, there is a
natural isomorphism between the Hilbert spaces $\ch$ and $\tch$.
\foot{This was done at finite value of the  regulator which controls the short
distance singularities.
In our analysis we shall keep the regulator finite throughout our
analysis, and set it to zero only at the very end.
For a coordinate independent description see ref.\NELSON.}
Let us denote the image of $|\tPh_{n,r}\r$ in $\ch$ under this isomorphism
also by $|\tPh_{n,r}\r$.
Then $|\Phi_{n,r}\r$ and $|\tPh_{n,r}\r$ are related by a linear
transformation of the form:
$$
|\tPh_{n,r}\r = \sum_s V^{[n]}_{sr} |\Phi_{n,s}\r
\eqn\etwosevena
$$
We can now expand $b_0^-|\Psi\r$ as,
$$
b_0^-|\Psi\r =\sum_r\psi'_r|\tPh_{2,r}\r
\eqn\etwosixa
$$
where $\psi_r$ and $\psi'_r$ are related by,
$$
\psi_r =\sum_s V^\TWO_{rs}\psi'_s
\eqn\etwosevenc
$$
Using eqs.\etwoone, \etwotwo, \etwoseven\ and \etwosixa\ we may express
$S(\Psi)$ and $\tilde S(\tilde\Psi)$ as,
$$
S(\Psi) =\sum_{N=2}^\infty {1\over N!} A^{(N)}_{r_1\ldots r_N}
\psi'_{r_1}\ldots \psi'_{r_N}
\eqn\etwoeight
$$
$$
\tS(\tPs) =\sum_{N=2}^\infty {1\over N!} \tA^\en_{r_1\ldots r_N}\tps_{r_1}
\ldots \tps_{r_N}
\eqn\etwonine
$$
where,
$$
A^\two_{r_1 r_2} =\l\tPh_{2,r_1}|Q_B c_0^-|\tPh_{2,r_2}\r
\eqn\etwoten
$$
$$
A^\en_{r_1\ldots r_N} = g^{N-2}\{(c_0^-\tPh_{2,r_1})\ldots
(c_0^-\tPh_{2,r_N})\} ~~{\rm for}~N\ge 3
\eqn\etwoeleven
$$
$$
\tA^\two_{r_1 r_2} =\l\tPh_{2,r_1}|\tqb c_0^-|\tPh_{2,r_2}\rpp
\eqn\etwotwelve
$$
$$
\tA^\en_{r_1\ldots r_N} = g^{N-2}\{(c_0^-\tPh_{2,r_1})\ldots
(c_0^-\tPh_{2,r_N})\}'' ~~{\rm for}~N\ge 3
\eqn\etwothirteen
$$
We now seek the following form of functional relationship between
$\psi'_r$ and $\tps_r$:
$$
\psi'_r \equiv (V^\TWO)^{-1}_{rs} \psi_s = \psi^\zero_r +
\sum_{N=1}^\infty
{1\over N!} S^\no_{r s_1\ldots s_N}\tps_{s_1}\ldots \tps_{s_N}
\eqn\etwofourteen
$$
which satisfies eq.\etwofive.
Here $\psi^\zero_r$ and $S^\no_{r s_1\ldots s_N}$ are some constants.
Without any loss of generality we can choose $S^\no$ to be symmetric in
its last $N$ indices.
Proof of background independence of string field theory now reduces to
showing the existence of appropriate $\psi^\zero_r$ and $S^\no_{r
s_1\ldots s_N}$ satisfying
eq.\etwofive.
In order that the term linear in $\tps$ on the left hand side of
eq.\etwofive\  vanishes,
$b_0^-|\Psi^\zero\r=\sum_r\psi^\zero_r|\tPh_{2,r}\r$ must be a solution of
the classical equations of motion
derived from the action $S(\Psi)$.

Let us define,
$$
\hp_r = \psi'_r -\psi^\zero_r, ~~~~~ b_0^-|\hat\Psi\r
=\sum_r\hat\psi_r|\tPh_{2,r}\r
\eqn\etwofifteen
$$
Using this, we may express eq.\etwoeight\ as,
$$\eqalign{
S(\Psi) & = S(\Psi^\zero +\hat\Psi)= S(\Psi^\zero) +\sum_{N=2}^\infty
{1\over N!} \hA^\en_{r_1\ldots
r_N} \hp_{r_1}\ldots \hp_{r_N}\cr
&\equiv S(\Psi^\zero) +\hS(\hP)\cr
}
\eqn\etwosixteen
$$
where,
$$\eqalign{
\hA^\en_{r_1\ldots r_N} &= \sum_{M=2}^\infty {g^{M+N-2}\over (M-2)!} \{
(\Psi^\zero)^{M-2} (c_0^-\tPh_{2,r_1})\ldots (c_0^-\tPh_{2,r_N})\}\cr
&= g^N \{(c_0^-\tPh_{2,r_1})\ldots (c_0^-\tPh_{2,r_N})\}'
{}~~~~~~~~~{\rm for}~N\ge 3 \cr
\hA^\two_{r_1 r_2} &= \l\tPh_{2,r_1}|Q_B c_0^- |\tPh_{2,r_2}\r
+\sum_{M=3}^\infty {g^{M-2}\over (M-2)!} \{(\Psi^\zero)^{M-2}
(c_0^-\tPh_{2,r_1})  (c_0^-\tPh_{2,r_2})\}\cr
&=\l\tPh_{2,r_1}|\hqb c_0^-|\tPh_{2,r_2}\r\cr
}
\eqn\etwoseventeen
$$
where,
$$
\{A_1\ldots A_N\}' = \sum_{M=2}^\infty {g^{M-2}\over
(M-2)!}\{(\Psi^\zero)^{M-2} A_1\ldots A_N\}
\eqn\etwoeighteen
$$
$$
\hqb b_0^-|A\r = Q_B b_0^-|A\r +\sum_{M=3}^\infty {g^{M-2}\over (M-2)!}
[(\Psi^\zero)^{M-2} A]
\eqn\etwonineteen
$$
and,
$$
[A_1\ldots A_N]' =\sum_{M=2}^\infty {g^{M-2}\over (M-2)!}
[(\Psi^\zero)^{M-2} A_1\ldots A_N]
\eqn\etwotwenty
$$
Detailed properties of $\hqb$, $[A_1\ldots A_N]'$ and $\{A_1\ldots A_N\}'$
have been analyzed in ref.\SENPOLY.
Identifying the constant in eq.\etwofive\ with $S(\Psi^\zero)$ we can
now express this equation as,
$$
\hS(\hP) = \tS(\tPs)
\eqn\etwotwentyone
$$
with eq.\etwofourteen\ taking the form:
$$
\hp_r = \sum_{N=1}^\infty {1\over N!} S^\no_{r s_1\ldots
s_N}\tps_{s_1}\ldots \tps_{s_N}
\eqn\etwotwentytwo
$$
Using eqs.\etwonine, \etwosixteen, \etwotwentyone\ and \etwotwentytwo\ we
get,
$$
\tA^\two_{r_1 r_2} = S^\zero_{ s_1r_1} S^\zero_{ s_2r_2} \hA^\two_{s_1
s_2}
\eqn\etwotwentythree
$$
$$\eqalign{
\tA^\thr_{r_1 r_2 r_3} =&\hA^\two_{s_1 s_2} (S^\zero_{s_1 r_1} S^\one_{s_2
r_2 r_3} + S^\zero_{s_1 r_2} S^\one_{s_2 r_1 r_3} + S^\zero_{s_1 r_3}
S^\one_{s_2 r_1 r_2})\cr
& + S^\zero_{ s_1r_1} S^\zero_{ s_2r_2} S^\zero_{ s_3r_3}\hA^\thr_{s_1 s_2
s_3}\cr
}
\eqn\etwotwentyfour
$$
and more generally,
$$\eqalign{
\tA^\en_{r_1\ldots r_N} = &\sum_{\{n, N_1,\ldots, N_n\}\atop n\ge 2,
N_i\ge 1, \sum N_i =N} {1\over n!\prod_i N_i!} \Big(\hA^{(n)}_{s_1\ldots
s_n} S^{(N_1-1)}_{s_1
r_1\ldots r_{N_1}} S^{(N_2-1)}_{s_2 r_{(N_1+1)}\ldots r_{(N_1+N_2)}}\ldots
S^{(N_n-1)}_{s_n r_{(N-N_n+1)}\ldots r_N}\cr
&+{\rm All ~~ permutations ~~ of}~~ r_1\ldots r_N\Big)\cr
}
\eqn\etwotwentyfiven
$$

We shall consider the case where CFT and CFT$''$ are related to each other
by an exactly marginal perturbation.
Also, in this section, we shall work to first order in the perturbation
parameter $\lambda$.
For definiteness, we take CFT to be a direct sum of a theory of some free
scalar fields (one of which is time-like) and an internal conformal field
theory; and take the marginal operator that relates CFT to CFT$''$ to be a
dimension $(1,1)$ operator $\vp$ in this internal conformal field theory.
It was shown in ref.\BACKGND\ that if the two dimensional action for
CFT$''$ is obtained by adding a term of the form $-\lambda\int d^2
z\vp(z,\bar z)$ to the two dimensional action of CFT, then $\Psi^\zero$ is
given by,
$$
b_0^-|\Psi^\zero\r = {\sqrt 2\lambda\over g} c_1 \bar c_1 |\vp\r
\eqn\etwotwentysix
$$
It was also shown that to order $\lambda$, there is an operator $\cs$
(called $S$ in ref.\BACKGND) which
acts on states annihilated by $b_0^-$ and $L_0^-=\tL_0^-$, and relates the
BRST charge $\tqb$ of CFTG$''$ to $\hqb$ defined in eq.\etwonineteen\
through the relation:
$$
\hqb = \cs \tqb \cs^{-1}
\eqn\etwotwentyseven
$$
Furthermore, $\cs$ preserves the inner product between states:
$$
\l \cs\Phi_1|c_0^-|\cs\Phi_2\r = \l\Phi_1 | c_0^-|\Phi_2\rpp
\eqn\etwotwentyeight
$$
where
$|\Phi_1\r$ and $|\Phi_2\r$ are any two states annihilated by $b_0^-$ and
$L_0^-$.
Taking the matrix elements of both sides of eq.\etwotwentyseven\ between
the states $\cs|\tPh_{2,r_1}\r$ and $c_0^-\cs|\tPh_{2,r_2}\r$, and using
eq.\etwotwentyeight, we get,
$$
\l \cs\tPh_{2,r_1}|c_0^-\hqb|\cs\tPh_{2,r_2}\r
=\l\tPh_{2,r_1}|c_0^-\tqb|\tPh_{2,r_2}\rpp
\eqn\etwotwentynine
$$
Thus, if we define $S^\zero_{rs}$ through the relation,
$$
\cs|\tPh_{2,r}\r =\sum_rS^\zero_{sr}|\tPh_{2,s}\r
\eqn\etwothirty
$$
then, using eqs.\etwotwelve\ and \etwoseventeen, eq.\etwotwentynine\ may
be written in the form:
$$
\hA^\two_{s_1 s_2}S^\zero_{s_1 r_1} S^\zero_{s_2 r_2} =\tA^\two_{r_1 r_2}
\eqn\etwothirtyone
$$
This is precisely eq.\etwotwentythree.
Thus we see that $S^\zero_{sr}$ defined in eq.\etwothirty\ provides a
solution to eq.\etwotwentythree.

In the next sections we shall show how to obtain solutions of
eqs.\etwotwentyfour\ and also discuss some general features of the other
equations \etwotwentyfiven.
For our analysis it will be more convenient to define $\bp_r$ as,
$$
\bp_r = (S^\zero)^{-1}_{rs}\hp_s
\eqn\etwothirtysix
$$
Then the action $\hS(\hP)$ given in eq.\etwosixteen\ may be expressed as,
$$
\hS(\hP)\equiv \bS(\bP) ={1\over 2}\tA^\two_{rs}\bp_r\bp_s+
\sum_{N=3}^\infty {1\over N!}\bA^\en_{r_1\ldots r_N} \bp_{r_1}\ldots
\bp_{r_N}
\eqn\etwothirtyseven
$$
where,
$$
\bA^\en_{r_1\ldots r_N} = S^\zero_{s_1 r_1}\ldots S^\zero_{s_Nr_N}
\hA^\en_{s_1\ldots s_N},~~~{\rm for}~N\ge 3
\eqn\etwothirtytwo
$$
The relation \etwotwentytwo\ now takes the form:
$$
\bp_r =\tps_r  + \sum_{N=2}^\infty {1\over N!} \bS^\no_{rs_1\ldots s_N}
\tps_{s_1} \ldots \tps_{s_N}
\eqn\etwothirtysevena
$$
where,
$$
\bS^\no_{r r_1\ldots r_N} = S^\no_{s r_1\ldots r_N} (S^\zero)^{-1}_{rs}
,~~~{\rm for}~N\ge 2
\eqn\etwothirtythree
$$
Eqs.\etwotwentyfour\ and \etwotwentyfiven\ may now be writtten as,
$$
\tA^\thr_{r_1 r_2 r_3} =\bA^\thr_{r_1 r_2 r_3} + (\tA^\two_{r_1
s_1}\bS^\one_{s_1 r_2 r_3} +\tA^\two_{r_2 s_1}\bS^\one_{s_1 r_1 r_3} +
\tA^\two_{r_3 s_1}\bS^\one_{s_1 r_1 r_2})
\eqn\etwothirtyfour
$$
and,
$$\eqalign{
\tA^\en_{r_1\ldots r_N} =&\bA^\en_{r_1\ldots r_N} + \sum_{n,M,
N_1,\ldots N_n \atop 2\le M<N, 1\le n\le M, N_i\ge 2, M -n+\sum N_i=N}
{1\over n! (M-n)!
\prod_i N_i!} \Big(\bA^\em_{r_1\ldots r_{M-n} s_1\ldots s_n}\cr
&\times \bS^{(N_1-1)}_{s_1 r_{(M-n+1)}\ldots r_{(M-n+N_1)}}
\bS^{(N_2-1)}_{s_2 r_{(M-n+N_1+1)}\ldots r_{(M-n+N_1+N_2)}} \ldots
\bS^{(N_n-1)}_{s_nr_{(N-N_n+1)}\ldots r_N}\cr
&+ {\rm ~~All~permutations~of~} r_1\ldots r_N\Big)\cr
}
\eqn\efournine
$$

Before we conclude this section, we shall write down the gauge symmetries
of the action $\bS(\bP)$.
The action $\hS(\hP)$ is known to have a gauge symmetry of the
form\SENPOLY\
$$
b_0^-\delta (|\hP\r) =\hqb b_0^-|\Lambda\r +\sum_{N=3}^\infty
{g^{N-2}\over (N-2)!} [\hP^{N-2}\Lambda]'
\eqn\etwothirtyeight
$$
If we define,
$$
b_0^-|\bP\r =\sum_r\bp_r|\tPh_{2,r}\r = \cs^{-1} b_0^-|\hP\r
\eqn\etwoforty
$$
then eqs.\etwothirtyeight\ and \etwotwentyseven\ gives,
$$
b_0^-\delta |\bP\r = \tqb b_0^- |\bar\Lambda\r +\sum_{N=3}^\infty
{g^{N-2}\over (N-2)!} \cs^{-1}[\hP^{N-2}\Lambda]'
\eqn\etwofortyone
$$
where,
$$
b_0^-|\bar\Lambda\r = \cs^{-1}b_0^-|\Lambda\r\equiv
\sum_r\bar\lambda_r|\tPh_{2,r}\r
\eqn\etwofortytwo
$$

Eq.\etwofortyone, together with the relations \etwoforty\ and
\etwofortytwo\ give the gauge symmetries of the action $\bS(\bP)$ given in
eq.\etwothirtyseven.
Note that the kinetic terms and the linearized gauge transformations have
the same form for $\tS(\tPs)$ and $\bS(\bP)$.
This fact will be useful to us in the later analysis.

\chapter{ANALYSIS OF CUBIC TERMS IN THE ACTION}

In this section we shall show that it is possible to choose suitable
$S^\one_{s r_1 r_2}$ (or, equivalently, $\bS^\one_{s r_1 r_2}$) so as to
satisfy eq.\etwotwentyfour\ (or, equivalently, eq.\etwothirtyfour).
We shall choose to work with the quantities $\bS^\one_{s r_1 r_2}$ and
eq.\etwothirtyfour, since the various equations we shall encounter during
this analysis take a more compact form in terms of these variables.
To begin with, we divide the basis states $\{|\tPh_{n,r}\r\}$ into three
sets: physical states $\{\tPh_{n,k_n}\r\}$, unphysical states
$\{|\tPh_{n,\alpha_n}\r\}$, and pure gauge states
$\{\tPh_{n,\alpha_{n-1}}\r\}$, satisfying the relations\BACKTWO:
$$\eqalign{
\tqb |\tPh_{n,k_n}\r =& 0\cr
\sum_{k_n} a_{k_n}|\tPh_{n,k_n}\r \ne& \tqb b_0^-|s\r~{\rm for~any
}~\{a_{k_n}\} ~{\rm and~any}~|s\r\cr
}
\eqn\ethreeone
$$
$$
\tqb \sum_{\alpha_n} a_{\alpha_n} |\tPh_{n,\alpha_n}\r \ne 0 ~{\rm
for~any}~\{a_{\alpha_n}\}
\eqn\ethreetwo
$$
$$
|\tPh_{n,\alpha_{n-1}}\r = \tqb |\tPh_{n-1,\alpha_{n-1}}\r
\eqn\ethreethree
$$
We shall further group the BRST invariant basis states
$\{|\tPh_{n,k_n}\r\}$ and $\{|\tPh_{n,\alpha_{n-1}}\r\}$ into a single
group, and denote this by $\{|\tPh_{n,\mu_n}\r\}$.
We shall now study eq.\etwothirtyfour\ for different choices of the
indices $r_1$, $r_2$ and $r_3$, and show that it is possible to choose
$\bS^\one_{s r r'}$ such that eq.\etwothirtyfour\ is satisfied for all possible
values of the indices $r_1$, $r_2$ and $r_3$.

\noindent {\bf Case I. All three indices correspond to unphysical states}

In this case, let us choose,
$$
|\tPh_{2,r_1}\r = |\tPh_{2,\alpha_2}\r, ~~~ |\tPh_{2,r_2}\r =
|\tPh_{2,\beta_2}\r, ~~~ |\tPh_{2,r_3}\r = |\tPh_{2,\gamma_2}\r
\eqn\ethreefour
$$
Since $\tA^\two_{rs}=\l\tPh_{2,r}|\tqb c_0^-|\tPh_{2,s}\r$ is
non-vanishing only when $|\tPh_{2,r}\r$ and $|\tPh_{2,s}\r$ are unphysical
states, the only non-vanishing components of $\tA^\two$ are of the form
$\tA^\two_{\at\bt}$.
We may now express eq.\etwothirtyfour\ as,
$$
\tA^\thr_{\at\bt\gt} =\bA^\thr_{\at\bt\gt}
+(\tA^\two_{\at\dt}\bS^\one_{\dt\bt\gt}
+\tA^\two_{\bt\dt}\bS^\one_{\dt\at\gt}
+\tA^\two_{\gt\dt}\bS^\one_{\dt\at\bt})
\eqn\ethreefive
$$
{}From eq.\ethreetwo\ and the definition of $\tA^\two$ it is clear that the
matrix $\tA^\two_{\at\bt}$ does not have any left or right eigenvector
with zero eigenvalue.
Hence it is an invertible matrix.\foot{ Although $\tA^\two_{\at\bt}$ is an
infinite dimensional matrix, it is block diagonal in the basis which are
chosen to be $\tL^+_0$ and the momentum eigenstates, with each block a
finite dimensional matrix.
Thus the results for finite dimensional matrices can be applied here.}
Let $M_{\at\bt}$ be this inverse.
Eq.\ethreefive\ may then be satisfied by choosing,
$$
\bS^\one_{\at\bt\gt} = {1\over 3}M_{\at\dt} (\tA^\thr_{\dt\bt\gt}
-\bA^\thr_{\dt\bt\gt} + L^\one_{\dt\bt\gt})
\eqn\ethreesix
$$
where $L^\one_{\dt\bt\gt}$ is an arbitrary tensor which is symmetric in
$\bt$ and
$\gt$ and satisfies:
$$
L^\one_{\at\bt\gt} +L^\one_{\bt\gt\at}+ L^\one_{\gt\at\bt} =0
\eqn\ethreeseven
$$
In particular, we may take $L^\one$ to be 0, but eq.\ethreeseven\
represents the most general solution of eq.\ethreefive.
As we shall see later in this section, $L^\one$ needs to be adjusted
appropriately in order to ensure that the coefficients $\bS^\one$ do not
have any discontinuity as a function of the momenta of various external
states.

\noindent{\bf Case II. One of the indices corresponds to a BRST invariant
state, others correspond to unphysical states}

Since eq.\etwothirtyfour\ is completely symmetric in the indices $r_1$,
$r_2$ and $r_3$, we may, without any loss of generality, choose:
$$
|\tPh_{2,r_1}\r = |\tPh_{2,\mu_2}\r, ~~~ |\tPh_{2,r_2}\r =
|\tPh_{2,\at}\r, ~~~ |\tPh_{2,r_3}\r = |\tPh_{2,\bt}\r
\eqn\ethreeeight
$$
Eq.\etwothirtyfour\ then takes the form:
$$
\tA^\thr_{\mt\at\bt} =\bA^\thr_{\mt\at\bt} + (\tA^\two_{\at\dt}
\bS^\one_{\dt\mt\bt} + \tA^\two_{\bt\dt} \bS^\one_{\dt\mt\at} )
\eqn\ethreenine
$$
This equation may be solved by choosing,
$$
\bS^\one_{\dt\mt\bt} = \bS^\one_{\dt\bt\mt} = {1\over 2} M_{\dt\at}
(\tA^\thr_{\at\mt\bt} -\bA^\thr_{\at\mt\bt} +L^{\prime\one}_{\at\bt\mt})
\eqn\ethreeten
$$
where $L^{\prime\one}$ is an arbitrary tensor, satisfying,
$$
L^{\prime\one}_{\at\bt\mt} + L^{\prime\one}_{\bt\at\mt} = 0
\eqn\ethreeeleven
$$

\noindent{\bf Case III. Two of the indices correspond to BRST invariant
states, the third one corresponds to an unphysical state}

In this case we choose,
$$
|\tPh_{2,r_1}\r = |\tPh_{2,\mt}\r, ~~~ |\tPh_{2,r_2}\r = |\tPh_{2,\nt}\r,
{}~~~ |\tPh_{2,r_3}\r = |\tPh_{2,\at}\r
\eqn\ethreetwelve
$$
Eq.\etwothirtyfour\ now takes the form:
$$
\tA^\thr_{\mt\nt\at} = \bA^\thr_{\mt\nt\at} + \tA^\two_{\at\dt}
\bS^\one_{\dt\mt\nt}
\eqn\ethreethirteen
$$
which can be satisfied by choosing,
$$
\bS^\one_{\dt\mt\nt} = M_{\dt\at} (\tA^\thr_{\at\mt\nt}
-\bA^\thr_{\at\mt\nt})
\eqn\ethreefourteen
$$

\noindent{\bf Case IV. All three indices correspond to BRST invariant
states}

In this case, we choose,
$$
|\tPh_{2,r_1}\r = |\tPh_{2,\mt}\r, ~~~ |\tPh_{2,r_2}\r = |\tPh_{2,\nt}\r,
{}~~~ |\tPh_{2,r_3}\r = |\tPh_{2,\tt}\r
\eqn\ethreefifteen
$$
and eq.\etwothirtyfour\ takes the form:
$$
\tA^\thr_{\mt\nt\tt} = \bA^\thr_{\mt\nt\tt}
\eqn\ethreesixteen
$$
Note that the coefficients $\bS^\one$ have dropped out of this equation.
Hence this equation cannot be satisfied by adjusting $\bS^\one$, it must be
satisfied identically.
We shall now show that this is indeed the case.
First let us consider the case where at least one of the states
$|\tPh_{2,\mt}\r$, $|\tPh_{2,\nt}\r$ and $|\tPh_{2,\tt}\r$ correspond to
a pure gauge state.
Without any loss of generality we can take this to be the state
$|\tPh_{2,\mt}\r$.
Thus we take,
$$
|\tPh_{2,\mt}\r = |\tPh_{2,\ao}\r = \tqb |\tPh_{1,\ao}\r
\eqn\ethreeseventeen
$$
We shall now show that for this choice of $|\tPh_{2,\mt}\r$, both, the
left and the right hand side of eq.\ethreesixteen\ vanish as a
consequence of gauge invariance.
This is intuitively obvious, since it involves on-shell three point
amplitudes involving external pure gauge states; but we shall go through
the proof in some detail, since it will be useful to derive a similar
result in the next section.
Let $\{\l\tPh^c_{n,r}|\}$ denote a basis of states conjugate to
$|\tPh_{n,r}\r$, satisfying,
$$
\l\tPh^c_{n,r}|\tPh_{n,s}\r'' = \delta_{rs}
\eqn\ethreeeighteennew
$$
Expanding the gauge transformation parameter $|\tilde\Lambda\r$ given in
eq.\etwofour\ as,
$$
b_0^-|\tilde\Lambda\r =\sum_{\ao}\tl_\ao |\tPh_{1,\ao}\r
\eqn\ethreenineteen
$$
we may express eq.\etwofour\ as,
$$
\delta\tps_r =\sum_{N=0}^\infty {1\over N!} \tB^\en_{r\ao s_1\ldots s_N}
\tps_{s_1}\ldots \tps_{s_N}\tilde\lambda_{\ao}
\eqn\ethreeseventeen
$$
where,
$$
\tB^\zero_{r\ao} =\l\tPh^c_{2,r}|\tqb |\tPh_{1,\ao}\r''
=\l\tPh^c_{2,r}|\tPh_{2,\ao}\r'' =\delta_{r\ao}
\eqn\ethreeeighteen
$$
and,
$$
\tB^\en_{r\ao s_1\ldots s_N} = g^N
\l\tPh^c_{2,r}|[(c_0^-\tPh_{2,s_1})\ldots (c_0^-\tPh_{2,s_N})
(c_0^-\tPh_{1,\ao})]''\r''~~{\rm for}~ N\ge 1
\eqn\ethreenineteen
$$
Invariance of the action \etwonine\ under the infinitesimal transformation
given in
eq.\ethreeseventeen\ gives,
$$
\tA^\two_{r_1 r_2}\tB^\zero_{r_2\ao} =0
\eqn\ethreetwenty
$$
$$
(\tA^\two_{r_1 s} \tB^\one_{s \ao r_2} + \tA^\two_{r_2 s} \tB^\one_{s\ao
r_1} ) +\tA^\thr_{r_1 r_2 s}\tB^\zero_{s\ao} = 0
\eqn\ethreetwentyone
$$
$$\eqalign{
& \Big(\tA^\two_{r_1 s}\tB^\two_{s\ao r_2 r_3} + \tA^\two_{r_2 s}
\tB^\two_{s\ao
r_3 r_1} +\tA^\two_{r_3 s} \tB^\two_{s\ao r_1 r_2})
+ \tA^\four_{s r_1 r_2 r_3}\tB^\zero_{s\ao} \cr
&~~~~~ + (\tA^\thr_{s r_1 r_2} \tB^\one_{s \ao r_3} + \tA^\thr_{s r_1 r_3}
\tB^\one _{s \ao r_2} + \tA^\thr_{s r_2 r_3} \tB^\one_{s \ao r_1}) = 0
}
\eqn\ethreetwentyonea
$$
etc.
Let us now choose the indices $r_1$ and $r_2$ in eq.\ethreetwentyone\ such
that,
$$
|\tPh_{2, r_1}\r = |\tPh_{2, \nt}\r, ~~~ |\tPh_{2, r_2}\r = |\tPh_{2,
\tt}\r
\eqn\ethreetwentytwo
$$
In this case $\tA^\two_{\nt s}$ and $\tA^\two_{\tt s}$
vanish since $\l\tPh_{2,\nt}|\tqb=0 =
\l\tPh_{2,\tt}|\tqb$.
Using eq.\ethreeeighteen\ we may now bring
eq.\ethreetwentyone\ into the form:
$$
\tA^\thr_{\nt\tt\ao}=0
\eqn\ethreetwentythree
$$
In an exactly similar way the gauge transformation \etwofortyone\ may be
expressed in component form as,
$$
\delta\bp_r =\sum_{N=0}^\infty{1\over N!} \bB^\en_{r\ao s_1\ldots
s_N}\bp_{s_1} \ldots \bp_{s_N} \bl_\ao
\eqn\ethreetwentyfour
$$
where,
$$\eqalign{
\bB^\zero_{r\ao} =&\l\tPh^c_{2,r}|\tqb|\tPh_{1,\ao}\r''
=\tB^\zero_{r\ao}\cr
\bB^\en_{r\ao s_1\ldots s_N} =&
g^N\l\tPh^c_{2,r}|\cs^{-1}[(c_0^-\cs|\tPh_{2, s_1}\r) \ldots
(c_0^-\cs|\tPh_{2, s_N}\r) (c_0^-\cs|\tPh_{1,\ao}\r)]'\r''\cr
}
\eqn\ethreetwentyfive
$$
The gauge invariance of the action $\bS(\bP)$ given in
eq.\etwothirtyseven\ under the transformation \ethreetwentyfour\ gives an
equation identical to eq.\ethreetwentyone\ with $\tB^\one$ and $\tA^\thr$
replaced by $\bB^\one$ and $\bA^\thr$ respectively.
Choosing the various external states as in eq.\ethreetwentytwo\ we get,
$$
\bA^\thr_{\nt\tt\ao}=0
\eqn\ethreetwentysix
$$

Thus we see that eq.\ethreesixteen\ is satisfied trivially when at least
one of the external states is of the pure gauge type.
Hence we need to verify this equation when all the external states are
physical.
In this case the left hand side of the equation is proportional to
the physical three point amplitude calculated in the string field theory
formulated around CFT$''$.
On the other hand, the right hand side of the equation is proportional to
the physical three point amplitude calculated from the action $\bS(\bP)$,
or, equivalently, $\hS(\hP)$, which is related to $\bS(\bP)$ by a linear
field redefinition.
The equality of these two amplitudes was proved in complete detail in
ref.\BACKTWO.
Hence we reach the conclusion that eq.\ethreesixteen\ is satisfied
identically, thereby proving that it is possible to satisfy
eq.\etwothirtyfour\ (and hence eq.\etwotwentyfour) by appropriately
adjusting the coefficients $\bS^\one$.

\noindent{\bf Continuity of $S^\zero$, $S^\one$ as functions of external
momenta}

The equations \ethreesix, \ethreeten\ and \ethreefourteen, which determine
the components $\bS^\one_{\at rs}$ for different choices of the external
states $r$, $s$, appear to be drastically different.
This may be the cause of some alarm, since, if we choose a basis of states
for a given momentum, and vary the momentum continuously, then some of the
unphysical states in the basis for generic values of the momentum may
become BRST invariant states for some special values of the momentum.
Eqs.\ethreesix, \ethreeten\ and \ethreefourteen\ would then seem to
indicate that the components of $\bS^\one$ change discontinuously at these
special values of momenta.
To be more concrete, let us consider the case when all the space-like flat
directions have been compactified and have discrete momenta/winding
number, so that the only continuous index is the momentum associated with
the time like directions.
Let us denote this by $k^0$.
If we consider the set of states in $\ch$ with a fixed value of $k^0$,
then, for generic $k^0$, all
the states have $\tL_0^+\ne 0$, and hence can be divided into pure gauge
and unphysical states.
In this case we can choose the basis of unphysical and pure gauge states
in such a way that as we vary $k^0$ continuously, the unphysical states
change smoothly into unphysical states, and pure gauge states change
smoothly into pure gauge states.
This continues till we reach some special values of $k^0$ (say $k^0_c$)
for which $\tL_0^+=0$ states  appear in the basis, so that the basis may
contain physical states.
In this case, the basis of states may be so chosen that as we approach
$k^0_c$ from a neighbouring value of $k^0$, some of the unphysical states
in the basis become physical, the rest remains unphysical; similarly some
of the pure gauge states in the basis become physical, the rest remains
pure gauge.\foot{ Note that the usual choice of gauge,
$b_0^+=0$, is not a good
gauge choice for a general analysis, since the basis of physical,
unphysical, and pure gauge states in this gauge collapses at $\tL^0_+=0$
($k^0=k^0_c$).
In particular, at $k^0 = k^0_c$, new unphysical states may appear which do
not satisfy the $b_0^+=0$ gauge condition.
In other words, if we choose a basis of unphysical and pure gauge states
for $k^0\ne k^0_c$, then it is not true that in the $k^0\to k^0_c$ limit
the set of unphysical (pure gauge) states divide themselves into
unphysical (pure gauge) and physical states.

There, is, however, one case in which $b_0^+=0$ is a good gauge choice;
this is when at $k^0=k^0_c$ the only $\tL_0^+=0$ state that appears is
a tachyonic physical state.
In this case, in the $\tL^0_+\to 0$ limit, an unphysical state gets
converted to a physical state, without any other rearrangement of the
basis.
The examples that will be considered in sect.4 are all of this type, hence
the $b_0^+= 0$ gauge choice in that section will not cause any problem in
our analysis.

This also indicates that in order to extend our analysis of $N$-point
vertices in sect.4 and that in ref.\BACKTWO\ beyond the tachyonic states,
it may be more convenient to choose a gauge other than $b_0^+=0$.}
As a result, the coefficients $\bS^\one$ determined from eqs.\ethreesix,
\ethreeten\ and \ethreefourteen\ may jump discontinuously as the momentum
$k^0$ associated with one of the external states approach $k^0_c$.
We shall now show that such a situation may be avoided by a judicious
choice of the quantities $L^\one$ and $L^{\prime\one}$ appearing in
eqs.\ethreesix\ and \ethreeten\ respectively.

Let us start with eq.\ethreesix\ and assume that the momentum associated
with the state $|\tPh_{2,\gt}\r$ approaches $k^0_c$, and also that in this
limit the state $|\tPh_{2,\gt}\r$ approaches a specific physical state
$|\tPh_{2, \lt^\zero}\r$.
In this case eq.\ethreesix\ gives a value of $\bS^\one_{\at\bt\lt^\zero}$.
If we want this value to agree with the answer given in eq.\ethreeten, we
must demand that,
$$
L^\one_{\dt\lt^\zero\bt} =L^\one_{\dt\bt\lt^\zero} ={1\over 2}
(\tA^\thr_{\dt\bt\lt^\zero} -\bA^\thr_{\dt\bt\lt^\zero}) +{3\over
2}L^{\prime\one}_{\dt\bt\lt^\zero}
\eqn\eadone
$$
On the other hand, if the momentum associated with the state
$|\tPh_{2,\at}\r$ approaches $k^0_c$, the matrix
$\l\tPh_{2,\at}| \tqb c_0^- |\tPh_{2,\dt}\r''$ acquires a zero eigenvalue
with eigenvector $v_\dt=\delta_{\dt\lt^\zero}$.
Thus the corresponding eigenvalue of $M_{\at\dt}$ has a pole.
In order that $\bS^\one$ given in eq.\ethreesix\ is finite in this limit,
we must have,
$$
L^\one_{\lt^\zero \at\bt} = -(\tA^\thr_{\lt^\zero\at\bt} -
\bA^\thr_{\lt^\zero\at\bt})
\eqn\eadtwo
$$
We may now ask if eqs.\eadone\ and \eadtwo\ are compatible with the
constraint \ethreeseven\ that $L^\one$ must satisfy.
{}From these two equations we get,
$$
L^\one_{\dt\lt^\zero\bt} +L^\one_{\bt\dt\lt^\zero}
+L^\one_{\lt^\zero\bt\dt} =3 (L^{\prime\one}_{\dt\bt\lt^\zero}+
L^{\prime\one}_{\bt\dt\lt^\zero})
\eqn\eadthree
$$
The right hand side of eq.\eadthree\ vanishes identically by
eq.\ethreeeleven.
Hence eqs.\eadone\ and \eadtwo\ are compatible with eq.\ethreeseven.

Next we consider eq.\ethreeten\ and take the limit when the momentum
associated with the state $|\tPh_{2,\bt}\r$ approaches the critical value
$k^0_c$.
Let us suppose that in this limit the state $|\tPh_{2,\bt}\r$ approaches
the physical state $|\tPh_{2,\lt^\zero}\r$.
In this case eq.\ethreeten\ gives us a value of
$\bS^\one_{\dt\mt\lt^\zero}$.
In order that this value agree with that obtained from eq.\ethreefourteen,
we must have,
$$
L^{\prime\one}_{\at\lt^\zero\mt}=\tA^\thr_{\at\lt^\zero\mt}
-\bA^\thr_{\at\lt^\zero \mt}
\eqn\eadfour
$$
On the other hand, if in eq.\ethreeten\ we take the momentum associated
with the state $|\tPh_{2,\dt}\r$ approach $k^0_c$, then, in order that
$\bS^\one$ approaches a finite value as $k^0\to k^0_c$, we must demand,
$$
L^{\prime\one}_{\lt^\zero \bt\mt} = -(\tA^\thr_{\lt^\zero \mt\bt} -
\bA^\thr_{\lt^\zero \mt\bt})
\eqn\eadfive
$$
We now see that eqs.\eadfour\ and \eadfive\ satisfy the constraint
\ethreeeleven:
$$
L^{\prime\one}_{\lt^\zero\at\mt} + L^{\prime\one}_{\at\lt^\zero\mt} = 0
\eqn\eadsix
$$
Finally, note that in eq.\ethreefourteen, if we let the momentum
associated with the state $|\tPh_{2,\dt}\r$ approach $k^\zero_c$, then, in
order that $\bS^\one_{\dt\mt\nt}$ is finite in this limit, we must have,
$$
\tA^\thr_{\lt^\zero\mt\nt} - \bA^\thr_{\lt^\zero\mt\nt} =0
\eqn\eadseven
$$
which is part of the consistency condition given in eq.\ethreesixteen.

Similar remark also holds for the components $S^\zero_{rs}$.
Writing $S^\zero_{rs}=1 +\lambda K^\zero_{rs} +\co(\lambda^2)$, we may
express eq.\etwotwentythree\ as,
$$
\tA^\two_{rs} -\hA^\two_{rs} =\lambda (K^\zero_{ur}\hA^\two_{us} +
K^\zero_{us} \hA^\two_{ur}) +\co(\lambda^2)
\eqn\eadseven
$$
whose solution is given by,
$$
\lambda K^\zero_{\at\bt} ={1\over 2} M_{\at\dt} (\tA^\two_{\dt\bt}
-\hA^\two_{\dt\bt} + L^\zero_{\dt\bt})
\eqn\eadeight
$$
$$
\lambda K^\zero_{\at\mt} = M_{\at\dt}(\tA^\two_{\dt\mt} -
\hA^\two_{\dt\mt})
\eqn\emextra
$$
where $L^\zero$ is a tensor satisfying,
$$
L^\zero_{\at\bt}+ L^\zero_{\bt\at} =0
\eqn\eadnine
$$
In addition, eq.\etwotwentyseven\ (which, combined with
eq.\etwotwentyeight\ is more restrictive than eq.\etwotwentythree) gives a
further constraint on $K^\zero$:
$$
\lambda K^\zero_{\kt\ao} =\l\tPh^c_{2,\kt}|(\hqb - \tqb)|\tPh_{1,\ao}\r''
\eqn\ekeone
$$
(Note that here $\tA^\two_{\dt\mt}=0$, but $\hA^\two_{\dt\mt}\sim\lambda$.)
As before, if the momentum $k^0$ associated with the state
$|\tPh_{2,\bt}\r$ approaches $k^0_c$, then, in order that eq.\eadeight\
smoothly approaches eq.\emextra, we must have,
$$
L^\zero_{\dt\lt^\zero} = \tA^\two_{\dt\lt^\zero} -\hA^\two_{\dt\lt^\zero}
\eqn\eadten
$$
On the other hand, if the momentum $k^0$ associated with the state
$|\tPh_{2,\at}\r$ approaches $k^0_c$, then, in order that the right hand
side of eq.\eadeight\ is finite in this limit, we must have,
$$
L^\zero_{\lt^\zero \bt} = -(\tA^\two_{\lt^\zero\bt}
-\hA^\two_{\lt^\zero\bt} )
\eqn\eadeleven
$$
{}From eqs.\eadten\ and \eadeleven\ we get,
$$
L^\zero_{\bt\lt^\zero} + L^\zero_{\lt^\zero\bt} =0
\eqn\eadtwelve
$$
as is required by eq.\eadnine.

Finally, if we take the momentum $k^0$ of the state
$|\tPh_{2,\at}\r$ in eq.\emextra\ to approach $k^0_c$, then, in order
that the right hand side of this equation is finite in this limit, we must
have,
$$
\tA^\two_{\lt^\zero\mt} -\hA^\two_{\lt^\zero\mt} =0
\eqn\eadthirteen
$$
This equation was shown to be satisfied identically in ref.\BACKGND.

These examples illustrate, how, by adjusting the parameters $L^\zero$,
$L^\one$ and $L^{\prime\one}$, we may maintain the continuity of the
solutions $S^\zero$ and $S^\one$ as function of the momenta $k^0$ of
various external states.

Note that in eq.\emextra, by taking the limit where the momentum $k^0$
associated with the state $|\tPh_{2,\at}\r$ approaches $k^0_c$, we can
evaluate $K^\zero_{\lt^\zero\mt}$.
(An independent expression for $K^\zero_{\lt^\zero\ao}$ is obtained from
eq.\ekeone, but it is easy to check that this agrees with the expression
for $K^\zero_{\lt^\zero\ao}$ obtained from eq.\emextra\ by taking
$k^0\to k^0_c$ limit and setting $|\tPh_{2,\mt}\r =
|\tPh_{2,\ao}\r$.)
On the other hand, an expression for $K^\zero_{\lt^\zero\kt}$ was found in
ref.\BACKTWO\ (eq.(3.60)) from the analysis of three and higher point
functions.
Thus we must ensure that these two results agree with each other.
A similar situation arises also in case of $\bS^\one_{\kt\mt\nt}$.
As we shall see in the next section, analysis of four and higher point
functions determines the quantities $\bS^\one_{\kt\mt\nt}$; hence we must
ensure that this value agrees with the value of
$\bS^\one_{\lt^\zero\mt\nt}$ obtained by taking the $k^0\to k^0_c$ limit
of $\bS^\one_{\dt\mt\nt}$, where $k^0$ is the momentum associated with the
state $|\tPh_{2,\dt}\r$.

A general argument showing that this must be the case may be given as
follows.
As we have seen, by adjusting $L^\one$, $L^{\prime\one}$ and $L^\zero$, we
can ensure that none of the components of $S^\zero$ and $S^\one$ has any
singularity as a function of the momenta of the external states.
Furthermore, the coefficients $\tA^\en$ and $A^\en$ appearing in
eqs.\etwotwentythree-\etwotwentyfiven\ are also smooth functions of the
momenta of the external states.
Thus if we can find a set of solutions $S^\en$ which satisfy
eqs.\etwotwentythree-\etwotwentyfiven\ for generic values of the external
momenta ($k^0\ne k^0_c$), and set $S^\en(k^0_c)=\lim_{k^0\to k^0_c}
S^\en(k^0)$, then this value of $S^\en(k^0_c)$ must satisfy
eqs.\etwotwentythree-\etwotwentyfiven.
Since $S^\zero_{\kt\lt}$ and $S^\one_{\kt\mt\nt}$ determined in
ref.\BACKTWO\ and sect.4 respectively give solutions to
eqs.\etwotwentythree-\etwotwentyfiven, they must be compatible with the
results obtained by taking the $k^0\to k^0_c$ limit of $S^\zero_{\dt\lt}$
and $S^\one_{\dt\mt\nt}$, where $k^0$ is the momentum associated with the
state $|\tPh_{2,\dt}\r$.

Another set of consistency conditions are obtained by starting from
eq.\ekeone\ and going to a limit when the state
$|\tPh_{2,\ao}\r=\tqb|\tPh_{1,\ao}\r$ approaches a physical state
$|\tPh_{2,m_2^\zero}\r$ after appropriate normalization.
This would happen if in this limit $|\tPh_{1,\ao}\r$ reduces to a physical
state so that $\tqb |\tPh_{1,\ao}\r$ vanishes in this limit unless it is
divided by an appropriate normalization factor before we take the limit.
A consistency condition is then obtained by demanding that $K^\zero_{\kt
m_2^\zero}$ found this way agrees with the answer given in eq.(3.60) of
ref.\BACKTWO.
The analysis of appendix A shows that this is indeed the case.
(Since physical states of ghost number 1 appear only in the zero momentum
sector, we need to study only the $k=0$ sector for this analysis.)

There is another related point that we must mention before concluding this
section.
In proving the equality of the physical amplitudes calculated from the
actions $\tS(\tPs)$ and $\bS(\bP)$, we had assumed in ref.\BACKTWO\ that
the physical states are of the form $c_1\bar c_1|\tilde V\r$, where
$|\tV\r$ is a dimension (1,1) primary state in CFT$''$.
While this is true for generic values of momenta (see refs.\THORN\GHMU\
and references therein) there is a physical state at zero momentum which
cannot be expressed in this form.
This is the zero momentum dilaton state, given by,
$$
|D\r = (c_1 c_{-1} -\bar c_1 \bar c_{-1})|0\r
\eqn\ethreetwentyseven
$$
Since we must show the equality of the two sides of eq.\ethreesixteen\ for
{\it all} states, we must also prove this for the case where one or more
of the external states correspond to the state $|D\r$ given in
eq.\ethreetwentyseven.
We have shown in appendix A that this is indeed what happens, thereby
completing the proof that eq.\ethreesixteen\ is satisfied for all values
of $\mt$, $\nt$ and $\tt$.

\chapter{ANALYSIS OF FOUR AND HIGHER POINT TERMS OF THE ACTION}

In this section we shall sketch the analysis of eq.\efournine\ for $N\ge
4$, but shall not give a complete proof of existence of solutions to these
equations.
Let us start with eq.\efournine\ for $N=4$.
Following the same analysis as in the previous section, one can show that
this equation can be satisfied by adjusting the coefficients $\bS^\two$
when at least one of the indices $r_i$ correspond to unphysical states.
In order to avoid repetition, we shall not go through the whole analysis,
but only consider a single case as an example.
Consider the case,
$$
|\tPh_{2,r_1}\r =|\tPh_{2,\mt}\r,~~|\tPh_{2,r_2}\r =|\tPh_{2,\at}\r, ~~
|\tPh_{2,r_3}\r = |\tPh_{2,\bt}\r, ~~ |\tPh_{2,r_4}\r = |\tPh_{2,\gt}\r
\eqn\efourone
$$
In this case eq.\efournine\ for $n=4$ is satisfied if we choose,
$$\eqalign{
\bS^\two_{\dt\bt\gt\mt} =& {1\over 3} M_{\dt\at} \big[
\tA^\four_{\at\bt\gt\mt} -\bA^\four_{\at\bt\gt\mt} - (\bA^\thr_{s\at\bt}
\bS^\one_{s\gt\mt} +\bA^\thr_{s\at\gt} \bS^\one_{s\bt\mt}
+\bA^\thr_{s\bt\gt} \bS^\one_{s\at\mt} \cr
&+\bA^\thr_{s\at\mt} \bS^\one_{s\bt\gt} +\bA^\thr_{s\bt\mt}
\bS^\one_{s\at\gt} +\bA^\thr_{s\gt\mt} \bS^\one_{s\at\bt}) \cr
& -\tA^\two_{ss'} (\bS^\one_{s\at\bt} \bS^\one_{s'\gt\mt}
+\bS^\one_{s\bt\gt} \bS^\one_{s'\at\mt} +\bS^\one_{s\at\gt}
\bS^\one_{s'\bt\mt}) + L^\two_{\at\bt\gt\mt}]\cr
}
\eqn\efourtwo
$$
where $L^\two_{\at\bt\gt\mt}$ is an arbitrary tensor which is symmetric in
$\bt$ and $\gt$ and satisfies the relation:
$$
L^\two_{\at\bt\gt\mt} +L^\two_{\bt\gt\at\mt} +L^\two_{\gt\at\bt\mt} =0
\eqn\efourtwoa
$$

Similar analysis can be done for other choices of indices.
The only possible problem comes from the case where all the indices $r_i$
correspond to BRST invariant states, since in this case the terms
involving $\bS^\two$ drop out of eq.\efournine\ for $N=4$.
More precisely, for the choice,
$$
|\tPh_{2,r_1}\r = |\tPh_{2,\mt}\r, ~~ |\tPh_{2,r_2}\r = |\tPh_{2,\nt}\r,
{}~~ |\tPh_{2, r_3}\r = |\tPh_{2,\tt}\r, ~~ |\tPh_{2, r_4}\r =
|\tPh_{2,\st}\r
\eqn\efourthree
$$
eq.\efournine\ for $N=4$ takes the form:
$$\eqalign{
\tA^\four_{\mt\nt\tt\st} =& \bA^\four_{\mt\nt\tt\st} + \tA^\two_{ss'}
(\bS^\one_{s\mt\nt} \bS^\one_{s'\tt\st} + \bS^\one_{s\mt\tt}
\bS^\one_{s'\nt\st} + \bS^\one_{s\mt\st} \bS^\one_{s'\nt\tt})\cr
&+ (\bA^\thr_{s\mt\nt} \bS^\one_{s\st\tt} + \bA^\thr_{s\mt\st}
\bS^\one_{s\nt\tt} +\bA^\thr_{s\mt\tt} \bS^\one_{s\nt\st} + \bA^\thr_{s\nt\tt}
\bS^\one_{s\mt\st} \cr
&+\bA^\thr_{s\nt\st}\bS^\one_{s\mt\tt}
+\bA^\thr_{s\tt\st} \bS^\one_{s\mt\nt})\cr
}
\eqn\efourfour
$$

In the last set of terms on the right hand side of eq.\efourfour, the
sum over $s$ can be broken up into sum over physical, unphysical, and
pure gauge states.
Of these, the sum over pure gauge states vanishes, since, as we have seen
in the last section, $\bA^\thr_{s\mt\nt}$ vanish for such states.
In the second set of terms on the right hand side of eq.\efourfour\ the sum
over $s$ and $s'$ can be restricted to unphysical states, since
$\tA_{ss'}$ vanishes for other set of states.
Using eq.\ethreefourteen, and the fact that $M_{\at\bt}$ is the inverse of
$\tA_{\at\bt}$, we may bring eq.\efourfour\ to the form:
$$\eqalign{
&\tA^\four_{\mt\nt\tt\st} -\tA^\thr_{\mt\nt\at} M_{\at\bt}
\tA^\thr_{\bt\tt\st} -\tA^\thr_{\mt\tt\at} M_{\at\bt}
\tA^\thr_{\bt\nt\st} -\tA^\thr_{\mt\st\at} M_{\at\bt}
\tA^\thr_{\bt\nt\tt}\cr
=& [\bA^\four_{\mt\nt\tt\st} -\bA^\thr_{\mt\nt\at} M_{\at\bt}
\bA^\thr_{\bt\tt\st} -\bA^\thr_{\mt\tt\at} M_{\at\bt}
\bA^\thr_{\bt\nt\st} -\bA^\thr_{\mt\st\at} M_{\at\bt}
\bA^\thr_{\bt\nt\tt}]\cr
&+ (\bA^\thr_{\kt\mt\nt}\bS^\one_{\kt\st\tt} +
\bA^\thr_{\kt\mt\st}\bS^\one_{\kt\nt\tt} +
\bA^\thr_{\kt\mt\tt}\bS^\one_{\kt\st\nt} +
\bA^\thr_{\kt\nt\tt}\bS^\one_{\kt\mt\st} \cr
& + \bA^\thr_{\kt\nt\st}\bS^\one_{\kt\mt\tt} +
\bA^\thr_{\kt\tt\st}\bS^\one_{\kt\mt\nt}) \cr
}
\eqn\efourfive
$$

We shall first show that the above equation can be satisfied
when one of the external states (say $|\tPh_{2,\mt}\r$) is of the form of
a pure gauge state.
Let us take $|\tPh_{2,\mt}\r = |\tPh_{2,\ao}\r = \tqb |\tPh_{1,\ao}\r
=\tB^\zero_{r\ao}|\tPh_{2,r}\r$.
In this case, in eq.\efourfive, all factors of $\tA^\en_{\mt r_1\ldots
r_{N-1}}$ and $\bA^\en_{\mt r_1\ldots r_{N-1}}$ are replaced by $\tA^\en_{r
r_1\ldots   r_{N-1}} \tB^\zero_{r\ao}$ and $\bA^\en_{r r_1\ldots
r_{N-1}}\tB^\zero_{r\ao}$ respectively.
The left hand side of eq.\efourfive\ then takes the form:
$$
(\tA^\four_{r\nt\tt\st} -\tA^\thr_{r\nt\at} M_{\at\bt}
\tA^\thr_{\bt\tt\st} -\tA^\thr_{r\tt\at} M_{\at\bt}
\tA^\thr_{\bt\nt\st} -\tA^\thr_{r\st\at} M_{\at\bt}
\tA^\thr_{\bt\nt\tt})\tB^\zero_{r\ao}
\eqn\efourthreea
$$
Choosing $r_1 = \nt$, $r_2=\at$ in eq.\ethreetwentyone\ we get,
$$
\tA^\two_{\at\bt}\tB^\one_{\bt\ao\nt} +\tA^\thr_{\nt\at
r}\tB^\zero_{r\ao} =0
\eqn\efourthreeb
$$
Using this and similar equations with $\nt$ replaced by $\tt$, $\st$, and
the relation $\tA^\two_{\at\bt} M_{\bt\gt}
=\delta_{\at\gt}$,
eq.\efourthreea\ may be brought into the form:
$$
\tA^\four_{r\nt\tt\st}\tB^\zero_{r\ao}
+\tA^\thr_{\bt\tt\st}\tB^\one_{\bt\ao\nt}
+\tA^\thr_{\bt\nt\st}\tB^\one_{\bt\ao\tt}
+\tA^\thr_{\bt\nt\tt}\tB^\one_{\bt\ao\st}
\eqn\efourthreed
$$
On the other hand, choosing $r_1=\nt$, $r_2=\tt$, $r_3=\st$ in
eq.\ethreetwentyonea, and using the relations
$\tA^\two_{r\tt}=\tA^\two_{r \st}=\tA^\two_{r\nt}=0$, we get,
$$
\tA^\four_{r\nt\st\tt}\tB^\zero_{r\ao}
+\tA^\thr_{r\tt\st}\tB^\one_{r\ao\nt}
+\tA^\thr_{r\nt\st}\tB^\one_{r\ao\tt}
+\tA^\thr_{r\tt\nt}\tB^\one_{r\ao\st} = 0
\eqn\efourthreee
$$
Comparing the left hand side of eq.\efourthreee\ with the expression
given in eq.\efourthreed\ we see that they differ from each other only in
the fact that the sum over $r$ in eq.\efourthreee\ runs over all states,
whereas the sum over $\bt$ in eq.\efourthreed\ runs over unphysical states
only.
Since $\tA^\thr_{r\tt\st}$ etc. vanish by eq.\ethreetwentythree\ if
$|\tPh_{2,r}\r$ is a pure gauge state, we see that the only extra terms on
the left hand side of eq.\efourthreee\ involves terms where
$|\tPh_{2,r}\r$ corresponds to a physical state.
In other words, using eq.\efourthreee, the expression \efourthreed\ may be
brought into the form:
$$
-\tA^\thr_{\kt\tt\st}\tB^\one_{\kt\ao\nt}
-\tA^\thr_{\kt\nt\st}\tB^\one_{\kt\ao\tt}
-\tA^\thr_{\kt\nt\tt}\tB^\one_{\kt\ao\st}
\eqn\efourthreef
$$
Using a similar analysis, the terms inside the square bracket on the right
hand side of eq.\efourfive\ may be brought into the form of
eq.\efourthreef\ with $\tA^\thr$ and $\tB^\one$ replaced by $\bA^\thr$ and
$\bB^\one$ respectively.
Finally, the first three terms inside the parantheses in the right hand
side of eq.\efourfive\
vanish by eq.\ethreetwentythree\ for the choice
$|\tPh_{2,\mt}\r=|\tPh_{2,\ao}\r$.
The equation \efourfive\ then takes the form:
$$\eqalign{
&\bA^\thr_{\kt\nt\tt}\bS^\one_{\kt\ao\st} +
\bA^\thr_{\kt\nt\st}\bS^\one_{\kt\ao\tt} +
\bA^\thr_{\kt\st\tt}\bS^\one_{\kt\ao\nt} \cr
=& \bB^\one_{\kt\ao\nt}\bA^\thr_{\kt\tt\st} +
\bB^\one_{\kt\ao\tt}\bA^\thr_{\kt\nt\st} +
\bB^\one_{\kt\ao\st}\bA^\thr_{\kt\tt\nt} \cr
-& \tB^\one_{\kt\ao\nt}\tA^\thr_{\kt\tt\st}
- \tB^\one_{\kt\ao\tt}\tA^\thr_{\kt\nt\st}
- \tB^\one_{\kt\ao\st}\tA^\thr_{\kt\tt\nt}\cr
}
\eqn\efourthreeg
$$
Using the consistency condition
$\bA^\thr_{\kt\tt\st}=\tA^\thr_{\kt\tt\st}$ we see that eq.\efourthreeg\
is satisfied if we choose,
$$
\bS^\one_{\kt\ao\st} =\bB^\one_{\kt\ao\st} - \tB^\one_{\kt\ao\st}
\eqn\efourthreeh
$$

Thus we see that eq.\efourfive\ can be satisfied by appropriately
adjusting $\bS^\one_{\kt\ao\st}$ if at least one of the external states is
pure gauge.
It now remains to analyze eq.\efourfive\ for the case where the states
$|\tPh_{2,\mt}\r$, $|\tPh_{2,\nt}\r$, $|\tPh_{2,\tt}\r$ and
$|\tPh_{2,\st}\r$ are all physical.
We consider two cases separately.
First we shall analyze the case where the external momenta are such that
no physical state can appear as intermediate state either in $s$, $t$ or
$u$ channel.
This, in fact, is the generic situation, since for generic values of
external momenta, the states allowed by momentum conservation as the
intermediate states in the $s$, $t$ and $u$ channel have non-zero $L_0^+$
eigenvalue.
In this case, $\bA^\thr_{\kt\mt\nt}$, $\bA^\thr_{\kt\mt\tt}$,
$\bA^\thr_{\kt\mt\st}$, $\bA^\thr_{\kt\nt\tt}$, $\bA^\thr_{\kt\nt\st}$,
$\bA^\thr_{\kt\st\tt}$ vanish for all $\kt$, and hence the last set of
terms on the right hand side of eq.\efourfive\ vanish.
Also, in this case, $M_{\at\bt}$, being the inverse of $\tA^\two_{\at\bt}$
in the subspace of unphysical states (which in this case coincides with
the subspace of states which are not pure gauge) may be interpreted as the
propagator of the gauge fixed theory, both, for the theory described by
the action $\bS(\bp)$ and the theory described by the action $\tS(\tps)$.
As a result, the left and the right hand side of eq.\efourfive\ are
proportional to the amputated four point Green's functions in the theories
described by the actions $\tS(\tPs)$ and $\bS(\bP)$ respectively, with
BRST invariant external states.
Thus in this case the problem reduces to showing that the on-shell four
point functions in the two theories are identical.

This is precisely the problem that we analyzed in ref.\BACKTWO, where we
proved that the on-shell amplitudes in these two theories with arbitrary
number of tachyonic external legs are indeed identical.\foot{In
ref.\BACKTWO\ we worked with the action $\hS(\hP)$ instead of $\bS(\bP)$,
but since they are related by a linear field redefinition, they have the
same on-shell S-matrix elements.}
Although the proof of this result could not be extended to the case of
arbitrary external states due to some technical difficulties, we believe
that the result does hold for arbitrary external states, particularly in
view of the fact that many of the S-matrix elements involving higher
excited states in string theory can be determined by studying the S-matrix
elements involving tachyonic states near the poles.

We now turn to the case of special values of momenta, which allow physical
external states to propagate in the $s$, $t$ or $u$ channel.
We shall assume, for simplicity, that this happens only in one channel,
say the $\mt\nt\to\st\tt$ channel, although our analysis can be easily
extended to the case where this happens in more than one channel
simultaneously.
Also, in order to avoid technical difficulties similar to the one that
arose in the analysis of ref.\BACKTWO\ for arbitrary external states, we
shall restrict ourselves to the case where all the external states are
tachyonic, and the physical states that appear in the intermediate channel
are also tachyonic.
In this case, among the last set of terms on the right hand side of
eq.\efourfive, only the terms involving $\bA^\thr_{\kt\mt\nt}$ and
$\bA^\thr_{\kt\st\tt}$ survive, and  this equation can be written as,
$$\eqalign{
&\tA^\four_{\mt\nt\tt\st} -\tA^\thr_{\mt\nt\at} M_{\at\bt}
\tA^\thr_{\bt\tt\st} -\tA^\thr_{\mt\tt\at} M_{\at\bt}
\tA^\thr_{\bt\nt\st} -\tA^\thr_{\mt\st\at} M_{\at\bt}
\tA^\thr_{\bt\nt\tt}\cr
& - [\bA^\four_{\mt\nt\tt\st} -\bA^\thr_{\mt\nt\at} M_{\at\bt}
\bA^\thr_{\bt\tt\st} -\bA^\thr_{\mt\tt\at} M_{\at\bt}
\bA^\thr_{\bt\nt\st} -\bA^\thr_{\mt\st\at} M_{\at\bt}
\bA^\thr_{\bt\nt\tt}]\cr
= & (\bA^\thr_{\kt\mt\nt}\bS^\one_{\kt\st\tt} +
\bA^\thr_{\kt\tt\st}\bS^\one_{\kt\mt\nt}) \cr
}
\eqn\efoursix
$$

It is shown in appendix B that if $\bA^\thr_{\mt\nt\kt}$ and
$\bA^\thr_{\st\tt\kt}$ vanish for all $k_2$, then the left hand side of
eq.\efoursix\ vanishes.
Thus the left hand side of eq.\efoursix\ must be of the form:
$$
\bA^\thr_{\mt\nt\kt}\Sigma^\one_{\kt\tt\st} +\bA^\thr_{\tt\st\kt}
\Sigma^\one_{\kt\mt\nt}
\eqn\efourseven
$$
for some tensor $\Sigma^\one$.
Thus eq.\efoursix\ may be satisfied by choosing,
$$
\bS^\one_{\kt\st\tt} = \Sigma^\one_{\kt\st\tt}
\eqn\efoureight
$$
Note that $\bS^\one_{\kt\st\tt}$, which was left undetermined during the
analysis of three point functions, is determined during the analysis of
four point functions.
The situation is analogous to the one encountered in
refs.\BACKGND\BACKTWO, where the components $S^\zero_{\kt\lt}$ were left
undetermined in the analysis of two point functions in ref.\BACKGND, but
were determined during the analysis of on-shell three point functions in
ref.\BACKTWO.

Let us now turn to the analysis of five and higher point functions.
The relevant equation to be satisfied is eq.\efournine.
The term involving $\bS^{(N-2)}$ on the right hand side of this equation
is of the form:
$$
(\tA^\two_{r_1 s}\bS^{(N-2)}_{s r_2\ldots r_N} + {\rm ~Permutations~of}~
r_1, \ldots r_N)
\eqn\efourten
$$
Generalizing the analysis for three and four point functions,
we see that if at least
one of the indices $r_i$ is unphysical, then eq.\efournine\ may be
satisfied by adjusting $\bS^{(N-2)}_{\at s_1\ldots s_{N-1}}$.
Thus the only non-trivial constraint comes when all the indices
$r_1,\ldots r_N$ correspond to BRST invariant states.
Let us take,
$$
|\tPh_{2, r_i}\r = |\tPh_{2,\mt^{(i)}}\r, ~~~~1\le i\le N
\eqn\efoureleven
$$
Then eq.\efournine\ takes the form:
$$\eqalign{
\tA^\en_{\mt^\one\ldots \mt^\en} = & \bA^\en_{\mt^\one\ldots \mt^\en} +
\sum_{n, M, N_1,\ldots N_n\atop 2\le M<N, 1\le n\le M, N_i\ge 2, M+\sum
N_i-n=N} {1\over n! (M-n)!
\prod_i N_i!}\cr
& \times \Big(\bA^\em_{\mt^\one\ldots \mt^{(M-n)} s_1\ldots s_n}
\bS^{(N_1-1)}_{s_1\mt^{(M-n+1)}\ldots \mt^{(M-n+N_1)}} \ldots
\bS^{(N_n-1)}_{s_n \mt^{(N-N_n+1)}\ldots \mt^\en}\cr
& +
{\rm ~~All~permutations~of}~\mt^\one\ldots \mt^\en\Big)\cr
}
\eqn\efourtwelve
$$

Let us first consider the case where the external momenta are such that no
physical state can propagate in the intermediate channel.
In this case, the sum over all the $s_i$'s in eq.\efourtwelve\ may be
taken to be over unphysical states only.
Since the analysis of eq.\efournine\ for a given value of $N$ determines
$\bS^{(N-2)}_{\at r_1\ldots r_{N-1}}$ in terms of the coefficients $\tA$,
$\bA$, $M_{\at\bt}$ and $\bS^\mo$ for $M<N$, we can eliminate all the
$\bS$'s appearing on the  right hand side of eq.\efourtwelve\ in terms of
$\bA$ and $M_{\at\bt}$.
The resulting equation is an expression involving the coefficients
$\tA^\em$, $\bA^\em$ and $M_{\at\bt}$.
We shall now give a general argument to show that this equation must be of
the form,
$$
\bG^\en_{\mt^\one\ldots\mt^\en} = \tG^\en_{\mt^\one\ldots\mt^\en}
\eqn\efourthirteen
$$
where $\tG$ is proportional to the amputated Greens function calculated
with the propagator $i M_{\at\bt}$ and vertices $ i\tA^\en_{r_1\ldots
r_N}$; for definiteness we choose the proportionality factor in such  a
way that $\tA^\en$ appears with coefficient unity in the expression for
$\tG^\en$.
Similarly, $\bG^\en$ is proportional to the amputated Green's function
calculated with the propagator $i M_{\at\bt}$ and vertices
$i\bA^\en_{r_1\ldots r_N}$.

Although it should be possible, in principle, to give a detailed
combinatoric proof of eq.\efourthirteen, we shall give an indirect
argument here to show that this is indeed the case.
{}From general arguments based on path integrals (or combinatoric analysis
of diagrams) one can show that if two theories are related by a field
redefinition, then they must have the same on-shell S-matrix elements.
Since the set of equations \efourtwelve\ are the only set of constraints
required for showing the equivalence of the theories described by the
actions $\tS(\tPs)$ and $\bS(\bP)$, they must include the condition that
the on-shell S-matrix elements in the two theories are identical.
This would happen only if the combinarotic factors work out correctly, so
that eq.\efourtwelve\ reduces to eq.\efourthirteen.

Once eq.\efourtwelve\ is expressed in the form of eq.\efourthirteen, we
see that the left and the right hand sides of this equation vanish
identically due to gauge invariance if any of the external states is of
the pure gauge type.
Thus we only need to consider the case when all the external states are
physical.
Again, the problem was analyzed in ref.\BACKTWO, where eq.\efourthirteen\
was proved to be true for all tachyonic external states.
The same technical problems that plague the analysis of four point
amplitude also prevents us from proving eq.\efourthirteen\ for a general
set of external states.
But the result is expected to hold for general external states also in
view of the fact that amplitudes involving tachyonic states in string
theory also contain most of the information about the on-shell amplitudes
involving higher excited states.

Finally we turn to the case when one or more intermediate channels admit
physical states propagating in them.
We would expect that in this case eq.\efourtwelve\ may be satisfied by
appropriate choice of the coefficients $\bS^{(M-1)}_{k_2 r_1\ldots r_M}$,
as in
the case of three and four point functions.
We shall not analyze the general case, but illustrate this through an
example.
Let us consider eq.\efourtwelve\ for $N=5$, and let us suppose that
physical ($L_0^+=0$) states can appear in the $\mt^\one\mt^\two\to
\mt^\thr\mt^\four\mt^\five$ channel.
We may now eliminate $\bS^\one_{\at rs}$ and $\bS^\two_{\at rst}$ by
eqs.\ethreesix, \ethreeten, \ethreefourteen\ and analogs of eq.\efourtwo.
The final result is of the form:
$$\eqalign{
&\tG^\five_{\mt^\one\ldots\mt^\five} -\bG^\five_{\mt^\one\ldots\mt^\five}
\cr
=&\bG^\four_{\mt^\thr\mt^\four\mt^\five
k_2}\bS^\one_{k_2\mt^\one\mt^\two} +\bA^\thr_{\mt^\one\mt^\two k_2}
(\bS^\two_{k_2\mt^\thr\mt^\four\mt^\five} - \bS^\one_{k_2\at\mt^\thr}
M_{\at\bt} \bA^\thr_{\bt\mt^\four\mt^\five} \cr
& -\bS^\one_{k_2\at\mt^\four}
M_{\at\bt} \bA^\thr_{\bt\mt^\thr\mt^\five}
-\bS^\one_{k_2\at\mt^\five}
M_{\at\bt} \bA^\thr_{\bt\mt^\thr\mt^\four}) \cr
}
\eqn\efourfourteen
$$
In writing down the above equation, we have used the gauge invariance of
three and four point amplitudes to eliminate terms involving
$\bS^\two_{\ao rst}$ and $\bS^\one_{\ao rs}$.
We now consider, as before, only those configurations for which the only
$\tL_0^+=0$ states which can
propagate in the intermediate state in the $\mt^\one\mt^\two\to
\mt^\thr\mt^\four\mt^\five$ channel are tachyonic.
In this case, by going through an analysis similar to the one discussed in
appendix B, one can see that the left hand side of eq.\efourfourteen\
vanishes if $\bA^\thr_{\mt^\one\mt^\two k_2}$ and
$\bG^\four_{\mt^\thr\mt^\four\mt^\five k_2}$ vanish.
Hence this has the form:
$$
\bA^\thr_{\mt^\one\mt^\two k_2}\Sigma^\two_{k_2\mt^\thr\mt^\four\mt^\five}
+ \bG^\four_{\mt^\thr\mt^\four\mt^\five
k_2}\Sigma^\one_{k_2\mt^\one\mt^\two}
\eqn\efourfifteen
$$
Note  that we have taken the coefficient of $\bG^\four$ in
eq.\efourfifteen\ to be the same tensor $\Sigma^\one$ as the one that
appeared in eq.\efourseven.
This is due to the fact that the coefficient of the $\bG^\four$ term is
expected to be determined solely by the structure of the states
$|\tPh_{2,\mt^\one}\r$, $|\tPh_{2,\mt^\two}\r$ and the propagator
$M_{\at\bt}$, and hence is expected to be equal to the coefficient of the
$\bA^\thr$ term on the left hand side of eq.\efoursix\ with appropriate
external indices.\foot{ This is similar to the result of ref.\BACKTWO\
where the coefficient $S^\zero_{k_2l_2}$ required to satisfy the equality
of on-shell $N$-point amplitudes in the two theories turned out to be
independent of $N$.}
Using eq.\efoureight\ we see that the term involving $\bG^\four$ in
eq.\efourfifteen\ is identical to the term involving $\bG^\four$ on the
right hand side of eq.\efourfourteen.
Thus eq.\efourfourteen\ may be satisfied by choosing,
$$\eqalign{
&(\bS^\two_{k_2\mt^\thr\mt^\four\mt^\five} - \bS^\one_{k_2\at\mt^\thr}
M_{\at\bt} \bA^\thr_{\bt\mt^\four\mt^\five}
-\bS^\one_{k_2\at\mt^\four}
M_{\at\bt} \bA^\thr_{\bt\mt^\thr\mt^\five} \cr
& -\bS^\one_{k_2\at\mt^\five}
M_{\at\bt} \bA^\thr_{\bt\mt^\thr\mt^\four} \cr
=& \Sigma^\two_{\kt\mt^\thr\mt^\four\mt^\five}\cr
}
\eqn\efourfifteen
$$
The coefficients $\bS^\two_{\kt\mt^\thr\mt^\four\mt^\five}$ and
$\bS^\one_{\kt\at\mt}$ had not been determined previously.
Hence we can choose these coefficients appropriately so as to satisfy
eq.\efourfifteen, and hence eq.\efourfourteen.

The analysis of this and the previous section leaves the coefficients
$\bS^\no_{\ao r_1\ldots r_{N}}$ completely undetermined.
This ambiguity in determining $\bS$ can be traced to the fact that a field
redefinition of the form
$$
b_0^-|\tPs\r\to b_0^-|\tPs\r +\lambda \Big(\tqb
b_0^-|f(\tPs)\r +\sum_{N=3}^\infty {g^{N-2}\over (N-2)!} [(|f(\tPs)\r)
\tPs^{N-2}]''\Big)
\eqn\efieldredef
$$
leaves the action $\tS(\tPs)$ invariant to order $\lambda$ for any choice
of $|f(\tPs)\r$.
This induces a transformation of the coefficients $\bS^\no_{s r_1\ldots
r_N}$ which is reflected in the ambiguity in determining the coefficients
$\bS^\no_{\ao r_1\ldots r_N}$.

Finally, note that if we want to evaluate the right hand side of
eqs.\etwotwentyfour, \etwotwentyfiven, using the values of $S^\no_{s_1
r_1\ldots r_N}$ determined in this and the previous sections, and compare
them with the left hand sides of these equations, we need to perform
infinite sum over the indices $s_i$.
These sums can be performed using the techniques of refs.\LPP\SENFIELD, as
was done in refs.\SYCL.

Thus we have demonstrated in this section, through some specific examples,
how the analysis of sect.3 might be extended to construct appropriate
field redefinitions which converts the full classical string field theory
action
$S(\Psi)$ formulated around the conformal field theory CFT to the string
field theory action $\tS(\tPs)$ formulated around the conformal field
theory CFT$''$.

\chapter{BACKGROUNDS SEPARATED BY FINITE DISTANCE AND RELATED BY MARGINAL
DEFORMATION}

So far we have considered backgrounds which are infinitesimally close to
each other.
In this section we shall show how to use the results that we have obtained
so far to relate backgrounds that are not necessarily close to each other,
but are still connected to each other by a series of marginal
deformations.
An intuitive undestanding of the resullts of this section may be obtained
by noting that if, to first order in the perturbation parameter $\lambda$,
the string field theories formulated around CFT and CFT$''$
are related by field redefinition, then the result is also expected to be
valid for finite $\lambda$, since we can deform CFT to CFT$''$ by marginal
deformation in infinitesimal steps.

To be more specific, let us now consider a one parameter family of allowed
backgrounds, such that any two neighbouring members of the family are
related to each other by a marginal perturbation.
Let $\tau$ denote the parameter labelling these backgrounds.
Then, given the conformal field theories at $\tau$ and $\tau+\delta \tau$,
the
two dimensional action of these two conformal field theories are related
by,
$$
\cs_{\tau+ \delta\tau} = \cs_\tau + f(\tau) \delta\tau \int d^2
z\vp(z,\bar z)
\eqn\efiveone
$$
where $\vp(z,\bar z)$ is a properly normalized dimension (1,1) primary
field.
$f(\tau)$ is some function of $\tau$ that can be set to 1 by appropriately
reparametrizing $\tau$.
For simplicity, we shall assume that this has been done.

Let $\{|\Phi_{n,r}(\tau)\r\}$ be an appropriate set of basis states in the
conformal field theory corresponding to a given value of $\tau$, and
$b_0^-|\Psi(\tau)\r =\sum_r \psi_r(\tau)|\Phi_{2,r}(\tau)\r$ be the string
field.
Identifying the conformal field theory corresponding to the point
$\tau$
to CFT$''$, that corresponding to the point $\tau+\delta\tau$ as CFT, and
using
eqs.\etwosevenc, \etwofourteen, we see that
$\psi_r(\tau)$ and $\psi_r(\tau+\delta \tau)$ are related by a functional
relationship of the form:
$$
\psi_s(\tau+\delta\tau) = V^\TWO_{sr}(\tau) (\psi^\zero_r(\tau)
+\sum_{N=1}^\infty
{1\over N!} S^{(N-1)}_{r s_1\ldots s_N}\psi_{s_1}(\tau) \ldots
\psi_{s_N}(\tau))
\eqn\efivetwo
$$
Since $\psi_s(\tau+\delta\tau)$ and $\psi_s(\tau)$ are expected to differ
by an amount of
order $\delta\tau$ (which was called $\lambda$ in the previous
sections), we may rewrite eq.\efivetwo\ as,
$$
\psi_s(\tau+\delta\tau) =\psi_s(\tau) +\delta\tau \sum_{N=0}^\infty
{1\over N!} T^{(N)}_{s r_1\ldots r_N}(\tau) \psi_{r_1}(\tau) \ldots
\psi_{r_N}(\tau)
\eqn\efivethree
$$
Note that we have included a $\tau$ dependence in $T^{(N)}$ since these
coefficients depend on the correlation functions in the conformal field
theory labelled by the parameter $\tau$.

Let $\tau_0$ be some fixed reference point in the $\tau$ space.
We shall now try to show that there is a field redefinition of the form:
$$
\psi_s(\tau) = \sum_{N=0}^\infty {1\over N!} R^\en_{s r_1\ldots
r_N}(\tau, \tau_0) \psi_{r_1}(\tau_0)\ldots \psi_{r_N}(\tau_0)
\eqn\efivefour
$$
which relates the string field theory actions formulated around the points
$\tau$ and $\tau_0$
even when $\tau$ and $\tau_0$ are not close to each other.
Using eqs.\efivefour\ we get,
$$
\psi_s(\tau+\delta\tau) -\psi_s(\tau) =\delta\tau \sum_{N=0}^\infty
{1\over N!} {d
R^\en_{s r_1\ldots r_N}(\tau, \tau_0)\over d\tau} \psi_{r_1}(\tau_0)
\ldots \psi_{r_N}(\tau_0)
\eqn\efivefive
$$
On the other hand, replacing $\psi_{r_i}(\tau)$ in the right hand side of
eq.\efivethree\ in terms of $\{\psi_{r_i}(\tau_0)\}$ given in
eq.\efivefour\ we get,
$$\eqalign{
&\psi_s(\tau+\delta\tau) -\psi_s(\tau)\cr
= & \delta\tau \sum_{N=0}^\infty {1\over N!}
T^\en_{s s_1\ldots s_N}(\tau) \Big(\sum_{M_1=0}^\infty {1\over M_1!}
R^{(M_1)}_{s_1 r_1\ldots r_{M_1}}\psi_{r_1}\ldots\psi_{r_{M_1}}\Big)\cr
& \ldots \Big(\sum_{M_N=0}^\infty {1\over M_N!}
R^{(M_N)}_{s_N r_{(M_1+\ldots M_{N-1}+1)} \ldots r_{(M_1+\ldots
M_N)}}\psi_{r_{(M_1+\ldots M_{N-1}+1)}}\ldots\psi_{r_{({M_1}+\ldots
{M_N})}}\Big)\cr
}
\eqn\efivesix
$$
Comparing eqs.\efivefive\ and \efivesix\ we get,
$$\eqalign{
{dR^\en_{s r_1\ldots r_N}(\tau, \tau_0)\over d\tau} = &
\sum_{n} {1\over n!} T^{(n)}_{s
s_1\ldots s_{n}}(\tau) \Big[\sum_{\{M_i\}, \sum M_i=N}
\Big(\prod_{i=1}^{n} {1\over M_i!}\Big) R^{(M_1)}_{s_1 r_1\ldots
r_{M_1}}\cr
& R^{(M_2)}_{s_2 r_{(M_1+1)} \ldots r_{(M_1+M_2)}} \ldots R^{(M_{n})}_{s_N
r_{(N-M_n+1)}\ldots r_{N}} +{\rm~~Permutations~of}~r_1,\ldots r_N\Big]\cr
}
\eqn\efiveseven
$$
These equations, together with the initial condition,
$$
R^\en_{s r_1\ldots r_N}(\tau, \tau_0)|_{\tau=\tau_0} =\delta_{N1}\delta_{s
r_1}
\eqn\efiveeight
$$
which follows from eq.\efivefour, determines $R^\en_{s r_1\ldots
r_N}(\tau, \tau_0)$ for general values of $\tau$ which are not necessarily
close to $\tau_0$.

In order to see how eq.\efiveseven\ may be used for practical computation,
let us consider a special case, $N=0$.
This gives,
$$
{dR^\zero_s (\tau,\tau_0)\over d\tau} =\sum_n {1\over n!} T^{(n)}_{s
s_1\ldots
s_n}(\tau) R^\zero_{s_1}(\tau, \tau_0) \ldots R^\zero_{s_n}(\tau, \tau_0)
\eqn\efivenine
$$
{}From eq.\efivefour\ we see that $R^\zero_s(\tau, \tau_0)$ gives
the value of $\psi_s(\tau)$ corresponding to the point
$\{\psi_r(\tau_0)=0\}$.
Since the point $\{\psi_r(\tau_0)=0\}$ describes the background
corresponding to the conformal field theory labelled by $\tau_0$, we see
that $R^\zero_s(\tau, \tau_0)$ gives us the coordinate of this particular
background in the coordinate system $\{\psi_s(\tau)\}$.
In other words,
$$
b_0^-|\Psi\r \equiv \sum_s R^\zero_s(\tau, \tau_0)
|\Phi_{2,s}(\tau)\r
\eqn\efiveninea
$$
is a solution of the classical equation of motion in string field theory
formulated around the point $\tau$ that describes the background labelled
by $\tau_0$.

We shall now derive an expression for $R^\zero_s(\tau, \tau_0)$ to order
$(\tau-\tau_0)^2$ by solving eq.\efivenine, and compare this with the
results of ref.\SYCL\ where a general algorithm was given for constructing
solutions of the classical equations of motion in string field theory to
all orders in the deformation parameter for (nearly) marginal
perturbation.
Since for $\tau\simeq\tau_0$, $R^\zero_s\propto (\tau-\tau_0)$, and since
we are interested only in the value of $R^\zero_s$ to order
$(\tau-\tau_0)^2$, we may express eq.\efivenine\ as,
$$
{dR^\zero_s(\tau, \tau_0)\over d\tau} = T^\zero_s(\tau) +T^\one_{sr}(\tau)
R^\zero_r(\tau) +\co\big((\tau-\tau_0)^2\big)
\eqn\efiveten
$$

Let us denote by $|\vp\r$ the dimension (1,1) primary state in the internal
conformal field theory representing the marginal deformation, and assume,
for simplicity, that $c_1\bar c_1|\vp\r$ is the only physical state
$|\tPh_{2,\kt}\r\equiv |\Phi_{2,\kt}(\tau)\r$ at zero momentum.
If we take $s=\kt$ in eq.\efiveten, we get,
$$
{dR^\zero_\kt(\tau, \tau_0)\over d\tau} = T^\zero_\kt(\tau) +T^\one_{\kt
r}(\tau) R^\zero_r(\tau)+\co\big((\tau-\tau_0)^2\big)
\eqn\efiveeleven
$$
Using eqs.\etwotwentysix, \efivetwo, \efivethree, and the  fact that
$V^\TWO_{sr}=\delta_{sr}+\co(\lambda)$, we get $T^\zero_\kt =\sqrt 2/g$,
$T^\zero_\ao = T^\zero_\bt =0$.
This gives,
$$
R^\zero_\kt (\tau, \tau_0) ={\sqrt 2\over g} (\tau-\tau_0)
+\co\Big((\tau-\tau_0)^2\Big)
\eqn\efivetwelve
$$
We shall now carry out a further reparametrization of $\tau$ such that the
right hand side of eq.\efivetwelve\ is exactly given by $\sqrt 2
(\tau-\tau_0)/g$.
Thus $R^\zero_\kt$ takes the form:
$$
R^\zero_\kt(\tau, \tau_0) ={\sqrt 2\over g} (\tau-\tau_0)
\eqn\efivethirteen
$$

Let us now concentrate on the components $R^\zero_\ao$ and $R^\zero_\bt$.
{}From eq.\efiveten, and the fact that $T^\zero_\ao=T^\zero_\bt=0$, we see
that $R^\zero_\ao$ and $R^\zero_\bt$ are of order $(\tau-\tau_0)^2$.
{}From eq.\efiveninea\ we see that $R^\zero_\ao$ can be set to zero to order
$(\tau-\tau_0)^2$ by a gauge transformation with the parameter
$-R^\zero_\ao |\Phi_{1,\ao}(\tau)\r$.
Thus $R^\zero_\bt$ are the only relevant components to be computed.
{}From eq.\efiveten\ we get,
$$
{dR^\zero_\bt\over d\tau} = T^\one_{\bt r}(\tau) R^\zero_r +
\co\Big((\tau-\tau_0)^2\Big)
\eqn\efivefourteen
$$
Note that the $\tau$ reparametrization needed to make eq.\efivethirteen\
an exact equation is of the form $\tau\to
\tau+\co\big((\tau-\tau_0)^2\big)$, and hence changes the right hand side
of eq.\efivefourteen\ only by a factor of order $(\tau-\tau_0)^2$.
Using eq.\efivetwelve\ and the fact that $R^\zero_\kt$ is the only
non-vanishing component of $R^\zero_r$ to order $(\tau-\tau_0)$, we get,
$$
R^\zero_\bt = {1\over\sqrt 2 g} T^\one_{\bt\kt}(\tau) (\tau-\tau_0)^2
+\co\Big((\tau-\tau_0)^3\Big)
\eqn\efivefifteen
$$
It thus remains to determine $T^\one_{\bt\kt}$.
It has been shown in appendix C that,
$$
T^\one_{\bt\kt} = -\sqrt 2\{(\tPh^c_{3,\bt}) (c_0^- c_1\bar c_1|\vp\r)
(c_0^- c_1\bar c_1|\vp\r)\}''
\eqn\efivefifteena
$$
Hence the classical solution to order $(\tau -\tau_0)^2$ may be written
as,
$$\eqalign{
b_0^-|\Psi\r = & {\sqrt 2\over g} \Big( (\tau-\tau_0) c_1\bar c_1 |\vp\r
-{1\over\sqrt 2} (\tau-\tau_0)^2 |\tPh_{2,\bt}\r \{(\tPh^c_{3,\bt}) (c_0^-
c_1\bar c_1|\vp\r) (c_0^- c_1\bar c_1|\vp\r)\}''\cr
&  + \co\Big((\tau
-\tau_0)^3\Big) \cr
}
\eqn\efivetwentyeight
$$
It can be easily verified that this agrees with the result computed in
ref.\SYCL\ to this order up to a gauge transformation and possible
reparametrization of $\tau$.\foot{ The notation for the choice of basis
given in ref.\SYCL\ is different from the one used here.
In order to translate the result of ref.\SYCL\ to the present notation, we
must make the replacement $\l\Phi^c_{n, \tilde k_n}|\to \l\Phi^c_{6-n,
k_{6-n}}|$, $\l\Phi^c_{n,\tilde\alpha_n}|\to\l\Phi^c_{6-n,
\alpha_{5-n}}|$, and $\l\Phi^c_{n,\tilde\alpha_{n-1}}|\to \l\Phi^c_{6-n,
\alpha_{6-n}}|$ for all $n$.}

\chapter{DISCUSSION}

In this paper we have shown that string field theories constructed around
two nearby conformal field theories that are related by marginal
perturbation are actually the same, in the sense that they are related to
each other by field redefinition.
Although the result has been proved in complete detail only for the quadratic
and cubic terms in the action, we have shown, by analyzing several special
cases, that the result is likely to be valid even for the quartic and
higher order terms in the action.
Finally, we have also shown that the result holds beyond leading
order in the perturbation parameter $\lambda$; as a result, string field
theories formulated around two different conformal field theories which
are not necessarily close to each other but are still connected to each
other by a series of marginal deformations, are also related to each
other through a field redefinition.
In this case, the appropriate field redefinition is found by solving an
infinite set of differential equations.

A question that naturally arises at this stage is whether it is possible
to formulate string field theory in a way where the background
independence is manifest.
For open string field theory such a formulation was given in reference
\CUBIC\ where it was shown that starting from a purely cubic action which
is independent of the background, one can obtain string field theories
around different backgrounds by shifting the field by a classical
solution.
The action expanded in terms of the shifted field has a background
dependent kinetic term, but a background independent interaction term.
Such a simple notion of background independence cannot, however, be
implemented in non-polynomial closed string field theory, since, by the
very nature of the non-polynomial interaction, a shift in the string field
will modify all the interaction vertices.

We would like to point out that the analysis of this paper may be applied
not only to study the equivalence of string field theories formulated
around two different backgrounds, but also to study the equivalence of
different formulations of string field theory around the same background.
The result of this paper tells us that if two string field theories have
the same kinetic terms, linearized gauge invariance, and on-shell S-matrix
elements (with special definition of `subtracted' S-matrix elements
when
physical states can propagate in the intermediate channel), then they can
be related to each other by field redefinition.

Finally, we would like to mention that the analysis of this paper (and
that of refs.\BACKGND\BACKTWO) have dealt with string field theory at the
classical level.
A complete quantum string field theory has been constructed\ZWIEBACH\
using the Batalin-Vilkovisky (BV) formalism\BATALIN, which requires adding
new terms to the string field theory action at the loop level.
A proof of background independence of the complete quantum string field
theory action will then involve showing that the actions of BV quantized
string field theories formulated around different backgrounds are related
to each other by appropriate field redefinitions, after taking into
account the change of path integral measure due to this field
redefinition.
We hope to return to this question in the near future.

\ack
I wish to thank A. Strominger and B. Zwiebach for discussion during early
stages of this work, and S. Mukherji for a critical reading of the
manuscript.

\Appendix{A}

In this appendix we shall try to verify eq.\ethreesixteen\ when one or
more of the external states $|\tPh_{2,\mt}\r$, $|\tPh_{2,\nt}\r$,
and $|\tPh_{2,\tt}\r$ correspond to a zero momentum dilaton
$$
|D\r = (c_1 c_{-1} -\bar c_1 \bc_{-1})|0\r
\eqn\eaone
$$
and the rest of the physical states are of the form $c_1\bc_1 |\tV\r$,
$|\tV\r$ being a dimension (1,1) primary state in CFT$''$.

Let us define,
$$
|s(k)\r = (c_1 c_{-1} -\bc_1\bc_{-1})|k\r + c_1\bc_1\eta_{\mu\nu}
\alpha^\mu_{-1}\bar\alpha^\nu_{-1} |k\r
\eqn\eappaone
$$
The starting point of our analysis will be the identity:
$$
\sk
= \Big(\eta_{\mu\nu} -{n_\mu k_\nu+ n_\nu k_\mu\over n.k}\Big) c_1\bar
c_1 \alpha^\mu_{-1}\bar\alpha^\nu_{-1}|k\r
-\tqb {n_\mu\over n.k} (c_1\alpha^\mu_{-1} -\bc_1\ba^\mu_{-1})|k\r
\eqn\eatwo
$$
where $k$ is an arbitrary momentum satisfying $k^2=0$, $\alpha^\mu_n$ are
the matter oscillators associated with
the flat directions, and $n^\mu$ is any vector.
This shows that $\sk$ has the form of $c_1\bc_1|\tV\r~~+$ pure gauge
state.
Hence for amplitudes involving such states, $\tA^\thr$ and $\bA^\thr$ have
the same values.
If we now take the $k\to 0$ limit of such amplitudes, then, if the $k\to
0$ limit is smooth, and is identical to the amplitude for $k=0$, this
would imply that $\tA^\thr$ and $\bA^\thr$ have the same values for
amplitudes involving external states of the form $(c_1
c_{-1}-\bc_1\bc_{-1})|0\r
+c_1\bc_1\eta_{\mu\nu} \alpha^\mu_{-1}\ba^\nu_{-1}|0\r$.
But the state $c_1\bc_1\eta_{\mu\nu}\alpha^\mu_{-1}\ba^\nu_{-1}|0\r$ has
the form $c_1\bc_1|\tV\r$, and hence, by  following the analysis of
ref.\BACKTWO\ one can see that $\tA^\thr$ and $\bA^\thr$ have the same
values for such external states.
This, in turn, would then imply that $\tA^\thr$ and $\bA^\thr$ are
identical even when one or more external physical states are of the form
$(c_1 c_{-1}-\bc_1\bc_{-1})|0\r$.

Thus it remains to show that the $k\to 0$ limit of the amplitudes
involving the state $\sk$ is smooth, both in $\tA^\thr$ and $\bA^\thr$,
and is identical to the corresponding amplitudes involving $|s(0)\r$.
For $\tA^\thr$ such amplitudes are obtained simply by taking the three
point correlation functions of appropriate vertex operators.
The $k\to 0$ limit of such an amplitude is manifestly smooth.
On the other hand, if we take $|\tPh_{2,\mt}\r=\sk$, then,
$$\eqalign{
\bA^\thr_{\mt\nt\tt}=&\hA^\thr_{rst} S^\zero_{r\mt} S^\zero_{s\nt}
S^\zero_{s\tt}\cr
=&\{(c_0^-\cs\sk)(c_0^-\cs|\tPh_{2,\nt}\r)(c_0^-\cs|\tPh_{2,\tt}\r)\}\cr
&+\{(c_0^-\sk)(c_0^-|\tPh_{2,\nt}\r)(c_0^-|\tPh_{2,\tt}\r)
(|\Psi^\zero\r)\} +\co(\lambda^2)\cr
}
\eqn\eathree
$$
The second term on the right hand side of eq.\eathree\ has well defined
$k\to 0$ limit and is identical to the $k=0$ answer.
The first term, in principle, can have a value whose $k\to 0$ limit does
not agree with the $k=0$ answer, since it involves the operator
$\cs=1+\lambda K+\co(\lambda^2)$.
$K$ is determined by the set of eqations discussed in
ref.\BACKGND\BACKTWO, and the algorithm given there was not manifestly
smooth in the $k\to 0$ limit, since it involved treating the physical,
unphysical, and pure gauge states on a different footing.
We shall now demonstrate explicitly that the $k\to 0$ limit of the
operator $K$ is actually smooth.
Once this is established, it would imply that the first term on the right
hand side of eq.\eathree\ also has smooth $k\to 0$ limit that agrees with
the $k=0$ result.

Thus we now need to show that the state $K\sk$ reduces, in the $k\to 0$
limit, to $K|s(0)\r$.
{}From the analysis of refs.\BACKGND\BACKTWO\ one can see that since $\sk$
is a physical state, only the components of the form
$\l\tPh_{2,\at}|c_0^-\tqb K\sk''$ and $\l\tPh_{3, k_3}|c_0^- K\sk''$ are
determined.
The equation determining $\l\tPh_{2,\at}|c_0^-\tqb K\sk''$ is,
$$
\lambda \l\tPh_{2,\at}|c_0^-\tqb K\sk'' = -\l\tPh_{2,\at}|c_0^-\Delta
Q_B\sk'' +\co(\lambda^2)
\eqn\eafour
$$
where,
$$\eqalign{
\l\Phi_1 |c_0^-\Delta Q_B |\Phi_2\r =& -\sqrt 2\lambda \{(c_0^-|\Phi_1\r)
(c_0^-c_1\bc_1 |\vp\r) (c_0^-|\Phi_2\r)\}\cr
&- \lambda\l\Phi_1| c_0^-\ointop_{|z|=\epsilon}\big(dz\bc(\bar z)\vp(z, \bar
z) -d\bar z c(z)\vp(z,\bar z)\big)|\Phi_2\r\cr
}
\eqn\eafive
$$
$|\vp\r$ being the same state that appeared in eq.\etwotwentysix, and
$\epsilon$ is the short distance cut-off used to define the correlation
functions in CFT$''$ in terms of those in CFT.
Since $\sk$ is a state built by the $\alpha^\mu_{-n}$, $\ba^\nu_{-n}$ and
the ghost oscillators on the state $|k\r$, it is clear from eqs.\eafive\
that the only states $\l\tPh_{2,\at}|$ for which the right hand side of
eq.\eafour\ does not vanish trivially are those built by the
$\alpha^\mu_n$, $\ba^\nu_n$ and the ghost oscillators on
$\l\vp|\otimes\l -k|$.
A simple analysis of BRST cohomology shows that the number of unphysical
states built on $\l\vp|\otimes \l -k|$ does have a smooth limit as $k\to
0$ since no new physical state appears at this value of the
momentum.\foot{ Note that we are taking the limit keeping $k^2=0$ all the
time.}
Thus it is possible to choose the basis of unphysical states
$\l\tPh_{2,\at}|$ for momentum $k$ in such a way that as $k\to 0$, these
states reduce to the unphysical states in the $k=0$ sector.
In this case the $k\to 0$ limit of the right hand side of eq.\eafour\
agrees with its $k=0$ value.
This, in turn, shows that the components $\l\tPh_{2,\at}|c_0^-\tqb K\sk''$
has smooth limit as $k\to 0$, and this limit agrees with the corresponding
expression with $\sk$ replaced by $|s(0)\r$.

Let us now analyze the components $\l\tPh_{3,k_3}|c_0^- K\sk''$.
Using eq.(3.60) of ref.\BACKTWO, and eq.\eatwo\ we see that,
$$\eqalign{
&\l\tPh_{3,k_3}|c_0^- K(\sk'' +\tqb {n^\mu\over n.k} (c_1\alpha^\mu_{-1}
-\bc_1\ba^\mu_{-1}) |k\r'')\cr
=& -2\ln({f_1'(0)\over\epsilon})\l\tPh_{3,k_3}|c_0^-\vp(1) (\eta_{\mu\nu}
-{n_\mu k_\nu + n_\nu k_\mu\over n.k})
c_1\bc_1\alpha^\mu_{-1}\bar\alpha^\nu_{-1}|k\r ''+\co(\lambda)\cr
}
\eqn\easix
$$
The only physical state $\l\tPh_{3, k_3}|$ for which eq.\easix\ is
non-trivial is $\l -k|\otimes \l\vp|c_{-1}\bc_{-1} c_0^+$.
But the right hand side of eq.\easix\ vanishes for such a state.
Using the equation $\lambda [K,\tqb]=\Delta Q_B$, we get,
$$
\lambda \l\tPh_{3, k_3}| c_0^- K\sk'' =- \l\tPh_{3, k_3}| c_0^-\Delta Q_B
{n^\mu\over n.k} (c_1 \alpha^\mu_{-1} -\bc_1\ba^\mu_{-1})|k\r'' +
\co(\lambda^2)
\eqn\easeven
$$

We shall now evaluate $\lambda\l\tPh_{3, k_3}|c_0^- K|s(0)\r''$ and compare
with the right hand side of eq.\easeven\ in the $k\to 0$ limit.
Using eq.(3.60) of ref.\BACKTWO\ we get,
$$\eqalign{
&\l\tPh_{3, k_3}| c_0^- K c_1\bc_1\alpha^\mu_{-1}\ba^\nu_{-1}
\eta_{\mu\nu} |0\r'' \cr
= & - 2\ln ({f_1'(0)\over \epsilon}) \l\vp|c_{-1}\bc_{-1} c_0^+ c_0^-
\vp(1) c_1\bc_1 \alpha^\mu_{-1}\ba^\nu_{-1}\eta_{\mu\nu} |0\r''
+\co(\lambda) \cr
= & \co(\lambda)\cr
}
\eqn\eaeight
$$
Using eqs.\eappaone, \easeven, and \eaeight\ we see that the $k\to 0$ limit
of $\l\tPh_{3, k_3}| c_0^- K \sk''$ matches $\l\tPh_{3, k_3}| c_0^- K
|s(0)\r''$ to order $\lambda$, if,
$$\eqalign{
&\l 0|\otimes \l\vp | c_{-1}\bc_{-1} c_0^+ c_0^- K (c_1 c_{-1}
-\bc_1\bc_{-1})|0\r'' \cr
=& - {1\over\lambda}\lim_{k\to 0}\l -k| \otimes \l\vp| c_{-1} \bc_{-1}
c_0^+ c_0^- \Delta
Q_B {n^\mu\over n.k} (c_1 \alpha^\mu_{-1} -\bc_1\ba^\mu_{-1}) |k\r'' +
\co(\lambda) \cr
=&-{f_1''(0)\over \sqrt 2 (f_1'(0))^2} \l 0|\otimes \l \vp| c_{-1}
\bc_{-1} c_0^+
c(f_2(1)) \bc(f_2(1)) \vp(f_2(1)) (c_1 - \bc_1)|0\r'' +\co(\lambda) \cr
}
\eqn\eanine
$$
We now note that the matrix element appearing on the left hand side of
eq.\eanine\ has not been determined previously.
Hence we can use eq.\eanine\ to define this matrix element.
This, in turn, guarantees that in the $k\to 0$ limit $K\sk$ reduces to the
state $K|s(0)\r$, and hence proves that eq.\ethreesixteen\ is satisfied
even when one or more of the external states correspond to the zero
momentum dilaton state $(c_1 c_{-1} -\bc_1 \bc_{-1})|0\r$.

The result of this appendix may also be interpreted in the following way.
Note that the pure gauge state,
$$
|\tPh_{2,\ao}\r \equiv \tqb {n_\mu\over n\cdot k} (c_1\alpha^\mu_{-1}
-\bar c_1\bar\alpha^\mu_{-1})|k\r
\eqn\eaextraa
$$
reduces to a physical state $|\tPh_{2,m_2^\zero}\r$ in the $k\to 0$ limit.
Our result shows that,
$$
\lim_{k\to 0}\l\tPh_{3, k_3}|c_0^- K|\tPh_{2,\ao}\r =\l\tPh_{3,k_3}|c_0^-
K|\tPh_{2,m_2^\zero}\r\big|_{k=0}
\eqn\eaextrab
$$
thereby showing that eq.\ekeone\ can be made compatible with eq.(3.60) of
ref.\BACKTWO.

\Appendix{B}

In this appendix we shall analyze the left hand side of eq.\efoursix\ and
show that it can be brought into the form of eq.\efourseven.
Using eq.\etwothirtytwo\ and \etwothirty\ and the fact\BACKGND\BACKTWO\
that when CFT and CFT$''$ differ by a marginal perturbation, then $\cs$
has the form:
$$
\cs = 1 +\lambda K +\co(\lambda^2)
\eqn\ebone
$$
we may express $\bA^\en_{\mt^\one\ldots \mt^\en}$ as,
$$
\bA^\en_{\mt^\one\ldots \mt^\en} =\hA^\en_{\mt^\one\ldots \mt^\en}
+\lambda \sum_{i=1}^N
K^\zero_{r\mu_2^{(i)}}\hA^\en_{r\mt^\one\ldots\mt^{(i-1)}
\mt^{(i+1)}  \mt^\en}
\eqn\ebtwo
$$
where $K^\zero_{rs}$ is defined through the relation,
$$
K|\tPh_{2,r}\r =\sum_s K^\zero_{sr}|\tPh_{2,s}\r
\eqn\ebthree
$$
i.e.
$$
K^\zero_{rs} =\l\tPh^c_{2,r}|K|\tPh_{2,s}\r''
\eqn\ebfour
$$
Let us define,
$$
\tilde\Delta_{rs} =\delta_{r\at}\delta_{s\bt} M_{\at\bt}
\eqn\ebfive
$$
The left hand side of eq.\efoursix\ may now be written as,
$$\eqalign{
L =&[\tA^\four_{\mt\nt\tt\st} -\tA^\thr_{\mt\nt t} \tilde\Delta _{ t s}
\tA^\thr_{ s\tt\st} -\tA^\thr_{\mt\tt t} \tilde\Delta _{ t s}
\tA^\thr_{ s\nt\st} -\tA^\thr_{\mt\st t} \tilde\Delta _{ t s}
\tA^\thr_{ s\nt\tt}]\cr
& - [\hA^\four_{\mt\nt\tt\st} -\hA^\thr_{\mt\nt t} \tilde\Delta _{ t s}
\hA^\thr_{ s\tt\st} -\hA^\thr_{\mt\tt t} \tilde\Delta _{ t s}
\hA^\thr_{ s\nt\st} -\hA^\thr_{\mt\st t} \tilde\Delta _{ t s}
\hA^\thr_{ s\nt\tt}]\cr
&-\lambda\big[ K^\zero_{r\mt} (\hA^\four_{r\nt\tt\st} -\hA^\thr_{r\nt
s}\tilde\Delta_{st}\hA^\thr_{t\tt\st}
- \hA^\thr_{r\tt
s}\tilde\Delta_{st}\hA^\thr_{t\nt\st} - \hA^\thr_{r\st
s}\tilde\Delta_{st}\hA^\thr_{t\tt\nt})\cr
& +{\rm~Cyclic~permutations~of~}\mt,\nt,\tt,\st\big] \cr
&+\lambda\big[(\hA^\thr_{\mt\nt
r}K^\zero_{rs}\tilde\Delta_{st}\hA^\thr_{t\tt\st} + \hA^\thr_{\mt\nt
s}\tilde\Delta_{st} K^\zero_{rt} \hA^\thr_{r\tt\st})\cr
&+{\rm ~Other~pairings~of}~\mt, \nt, \tt, \st\big] +\co(\lambda^2)\cr
}
\eqn\ebsix
$$
We assume that the only physical states that can appear in the
$\mt\nt\to\tt\st$ channel are tachyonic.
This, in turn, shows that if $k$ is the momentum flowing in this channel,
then the corresponding vertex operator $e^{ik.X}$ has dimension $h>0$.
As a result, the only possible $\tL^+_0=0$ states propagating in this
channel are of the form $c_1\bc_1 e^{ik.X(0)}|\tV\r$, where $|\tV\r$ is a
dimension $(1-h, 1-h)$ primary state in the internal conformal field
theory.
In other words, the only possible $\tL^0_+=0$ states propagating in this
channel are the tachyonic physical states.

This, in turn, implies that if $\tA^\thr_{\mt\nt\kt}$ and
$\tA^\thr_{\tt\st\kt}$
vanish for all $k_2$, then the only possible intermediate states
$|\tPh_{2,t}\r$, $|\tPh_{2,s}\r$
appearing in eq.\ebsix\ are the $\tL^+_0\ne 0$ states.
(We have already assumed that the momenta flowing in the $u$ and $t$
channels are such that they do not allow $\tL^+_0=0$ states to appear as
intermediate states in these channels).
We shall now show that the right hand side of eq.\ebsix\ vanishes in this
case.
In the $\tL^+_0\ne 0$ sector, a basis of unphysical states can be chosen all
of which are annihilated by $b_0^+$ (see, for example,
ref.\SIEGELBOOK).
This gives rise to the standard expression for the propagator
$\tilde\Delta_{rs}$:
$$
\tilde\Delta_{rs}
=\l\tPh^c_{2,r}|b_0^-b_0^+(\tL_0^+)^{-1}|\tPh^c_{2,s}\r''
\eqn\ebsixa
$$
Computation of the terms inside the first square bracket in eq.\ebsix\
then reduces to the standard computation of a four point function in
string field theory defined by the action $\tS(\tPs)$ with physical
external states.
Standard manipulations given in ref.\BACKTWO\ then shows that this
amplitude is identical to the corresponding amplitude calculated with
the action $\hS(\hP)$.
In the analysis of ref.\BACKTWO\ the kinetic term of $\hS(\hP)$ was split
into two
pieces:
$$
\l\hP|\hqb b_0^-|\hP\r =\l\hP|Q_B b_0^-|\hP\r +\l\hP|(\hqb - Q_B)
b_0^-|\hP\r
\eqn\ebseven
$$
The inverse of the first term on the right hand side of eq.\ebseven\ in
the $b_0^+=0$ gauge, i.e.,
$$
\Delta_{rs}=\l\Phi^{\prime c}_{2,r}|b_0^-b_0^+ (L_0^+)^{-1}|\Phi^{\prime
c}_{2,s}\r
\eqn\ebeight
$$
was taken as the propagator, whereas the second term on the right hand
side of eq.\ebseven\ was taken as the interaction term.
Here $\{\l\Phi^{\prime c}_{2,r}|\}$ is the basis of states conjugate to
$\{|\tilde \Phi_{2,r}\r\}$ with respect to the inner product $\l |\r$,
i.e.,
$$
\l\Phi^{\prime c}_{2,r}|\tilde\Phi_{2,s}\r =\delta_{rs}
\eqn\ebeighta
$$
Thus, by repeating the analysis  of ref.\BACKTWO\ one can show
that the contribution from the set of terms inside the first square
bracket on the right hand side of eq.\ebsix\ takes the form:
$$\eqalign{
& S^\zero_{ r\mt} S^\zero_{s\nt} S^\zero_{t\tt } S^\zero_{u\st}
\big\{\hA^\four_{rstu} - (\Delta_{r't'} -\lambda \Delta_{r'u'}\Lambda_{u's'}
\Delta_{s't'})\cr
&\times (\hA^\thr_{r' r s}\hA^\thr_{t'tu} +\hA^\thr_{r'rt}\hA^\thr_{t'su}
+\hA^\thr_{r' r u}\hA^\thr_{t' st})\big\} +\co(\lambda^2)\cr
}
\eqn\ebnine
$$
where,
$$
\lambda\Lambda_{rs}=\l\tPh_{2,r}|(\hqb - Q_B) c_0^- |\tPh_{2,s}\r
\eqn\ebten
$$
Let us define,
$$
\lambda\Omega_{rs} = \tilde\Delta_{rs} -\Delta_{rs}
\eqn\ebeleven
$$
Using eqs.\etwothirty, \ebone, \ebfour, \ebsix,
and that the first set of terms inside the square bracket in eq.\ebsix\
can be expressed as eq.\ebnine, we get,
$$\eqalign{
L =&\lambda\big\{ \hA^\thr_{\mt\nt r}\hA^\thr_{\tt\st s}
+ \hA^\thr_{\mt\st r}\hA^\thr_{\tt\nt s} + \hA^\thr_{\mt\tt
r}\hA^\thr_{\nt\st s} \big\}\cr
&\times \big\{ \Omega_{rs} +\Delta_{rr'}\Lambda_{r's'}\Delta_{s's} +
K^\zero_{rr'} \tilde\Delta_{r's} +\tilde\Delta_{rr'} K^\zero_{sr'}\big\}
+\co(\lambda^2) \cr
}
\eqn\ebthirteen
$$
We now note that,
$$\eqalign{
\lambda\Omega_{rs} =&\l\tPh^c_{2,r}|b_0^- b_0^+
(\tL^+_0)^{-1}|\tPh^c_{2,s}\r''
-\l\Phi^{\prime c}_{2,r}| b_0^- b_0^+ (L_0^+)^{-1}|\Phi^{\prime c}_{2,s}\r
\cr
=& \l\tPh^c_{2,r}|b_0^- b_0^+
\big((\tL^+_0)^{-1} - (L_0^+)^{-1}\big)|\tPh^c_{2,s}\r'' \cr
& + \l\tPh^c_{2,r}|b_0^- b_0^+
(L^+_0)^{-1}|\tPh^c_{2,s}\r''  - \l\tPh^c_{2,r}|b_0^- b_0^+
(L^+_0)^{-1}|\tPh^c_{2,s}\r\cr
& - \l\delta\Phi^c_{2,r}|b_0^- b_0^+
(L^+_0)^{-1}|\tPh^c_{2,s}\r
-\l\tPh^c_{2,r}|b_0^- b_0^+
(L^+_0)^{-1}|\delta\Phi^c_{2,s}\r +\co(\lambda^2)\cr
}
\eqn\ebthirteena
$$
where,
$$
\l\delta\Phi^c_{2,r}| =\l\Phi^{\prime c}_{2,r}| -\l\tPh^c_{2,r}|
\eqn\ebthirteenb
$$
{}From eqs.\ethreeeighteennew, \ebeighta\ and \ebthirteenb, we see that,
$$
\l\delta\Phi^c_{2,r}|\tPh_{2,s}\r =\l\tPh^c_{2,r}|\tPh_{2,s}\r''
-\l\tPh^c_{2,r}|\tPh_{2,s}\r
\eqn\ebthirteenc
$$
Since $\{|\tPh_{2,s}\r\}$ forms a complete basis of states,
eq.\ebthirteenc\ is valid with $|\tPh_{2,s}\r$ replaced by any state
$|A\r$.
Choosing $|A\r = b_0^- b_0^+ (L_0^+)^{-1}|\tPh^c_{2,s}\r$, we see that the
second, third, and the fourth terms on the right hand side of
eq.\ebthirteena\ cancel.
On the other hand, the first term on the right hand side of
eq.\ebthirteena\ may be expressed as,
$$\eqalign{
& -\l\tPh^c_{2,r}|b_0^-b_0^+ (\tL^+_0)^{-1} (\tL^+_0 - L^+_0) (\tL^+_0)^{-1}
|\tPh^c_{2,s}\r'' +\co(\lambda^2)\cr
=& -\lambda\Delta_{rr'} P_{r's'} \Delta_{s's} +\co(\lambda^2)\cr
}
\eqn\ebthirteend
$$
where,
$$
\lambda P_{r's'} =\l\tPh_{2,r'}| (\tqb - Q_B) c_0^- |\tPh_{2, s'}\r
\eqn\ebfifteen
$$
Finally, since $b_0^- b_0^+ (L_0^+)^{-1}$ is hermitian with respect to the
inner product $\l |\r$, using eq.\ebthirteenc\
the last term on the right hand side of eq.\ebthirteena\ may
be written as,
$$\eqalign{
& -\l\delta\Phi^c_{2,s}| b_0^- b_0^+ (L_0^+)^{-1}|\tPh^c_{2,r}\r\cr
=& -\l\tPh^c_{2,s}|b_0^-b_0^+ (L_0^+)^{-1}|\tPh^c_{2,r}\r''
+\l\tPh^c_{2,s}|b_0^-b_0^+ (L_0^+)^{-1}|\tPh^c_{2,r}\r +\co(\lambda^2)\cr
}
\eqn\ebthirteenf
$$
Eq.\ebthirteena\ may now be written as,
$$\eqalign{
\lambda\Omega_{rs} = & -\lambda\Delta_{rr'} P_{r's'}\Delta_{s' s}
-\l\tPh^c_{2,s}| b_0^- b_0^+ (L_0^+)^{-1}|\tPh^c_{2,r}\r''\cr
& + \l\tPh^c_{2,s}| b_0^- b_0^+ (L_0^+)^{-1}|\tPh^c_{2,r}\r
+\co(\lambda^2) \cr
}
\eqn\ebfourteen
$$

Next we use the relation,
$$
\tilde \Delta_{\bt s} \l\tPh_{2,s}| \tqb c_0^- |\tPh_{2,\at}\r
=\delta_{\at\bt}
\eqn\ebsixteen
$$
$$
\l\tPh_{2,\bt}| \tqb c_0^- |\tPh_{2, s}\r \tilde\Delta_{s\at}
=\delta_{\at\bt}
\eqn\ebseventeen
$$
to write,
$$
K^\zero_{\at r'}\tilde\Delta_{r'\bt} +\tilde\Delta_{\at r'} K^\zero_{\bt
r'}
=\tilde\Delta_{\at s} R_{st} \tilde\Delta_{t\bt}
\eqn\ebeighteen
$$
where,
$$
R_{st} =\l\tPh_{2,s}| \tqb c_0^- K|\tPh_{2,t}\r'' +
\l\tPh_{2,t}| \tqb c_0^- K|\tPh_{2,s}\r''
\eqn\ebnineteen
$$
Using the relation\BACKGND\BACKTWO,
$$
\lambda [\tqb, K] = -\Delta Q_B = -\hqb +\tilde Q_B
\eqn\ebtwenty
$$
we may express eq.\ebnineteen\ as,
$$\eqalign{
\lambda R_{st} =& \l\tPh_{2, s}|(\tqb -\hqb) c_0^- |\tPh_{2,t}\r''\cr
& - \l\tPh_{2, s}| c_0^- K \tqb |\tPh_{2,t}\r''
+ \l\tPh_{2, t}| \tqb c_0^- K |\tPh_{2,s}\r''\cr
}
\eqn\ebtwentyone
$$
We now note that in eq.\ebthirteen\ the sum over $r$ and $s$ run over
unphysical states only, since $\hA^\thr_{\mt\nt r}$ etc. are taken to be
non-vanishing only  for such states.\foot{ Actually, it is
$\tA^\thr_{\mt\nt r}$ etc. which vanish for $r\ne\at$, but this implies
that $\hA^\thr_{\mt\nt r}$ is of order $\lambda$ for $r\ne\at$.}
Taking $r=\at$ and $s=\bt$, and using eqs.\ebten, \ebfourteen, \ebfifteen,
\ebeighteen, and \ebtwentyone, we get,
$$\eqalign{
L =& \big (\hA^\thr_{\mt\nt\at} \hA^\thr_{\tt\st\bt}
+\hA^\thr_{\mt\st\at} \hA^\thr_{\tt\nt\bt}
+\hA^\thr_{\mt\tt\at} \hA^\thr_{\nt\st\bt}\big)\cr
&\times \big\{ -\l\tPh^c_{2,\bt}| b_0^-b_0^+ (L_0^+)^{-1}
|\tPh^c_{2,\at}\r''
+ \l\tPh^c_{2,\bt}| b_0^-b_0^+ (L_0^+)^{-1}
|\tPh^c_{2,\at}\r \cr
& -\lambda\tilde\Delta_{\at s} \l\tPh_{2, s}| c_0^- K \tqb |\tPh_{2, t}\r''
\tilde\Delta_{t\bt}
+ \lambda\tilde\Delta_{\bt t} \l\tPh_{2, t}|  \tqb c_0^- K |\tPh_{2, s}\r''
\tilde\Delta_{s\at} \big\} +\co(\lambda^2)\cr
}
\eqn\ebtwentytwo
$$
Summing over a complete set of states, the last two terms in the curly
bracket may be expressed as,
$$
\lambda\l\tPh^c_{2,\at}| b_0^- b_0^+ (\tL_0^+)^{-1} c_0^- K
b_0^-|\tPh^c_{2,\bt}\r ''
+ \lambda\l\tPh^c_{2,\bt}| K b_0^- b_0^+ (\tL_0^+)^{-1}
|\tPh^c_{2,\at}\r ''
\eqn\ebtwentythree
$$
Using eqs.\etwotwentyeight\ and \ebone, and that $b_0^- b_0^+
(\tL_0^+)^{-1}$ is hermitian with respect to the inner product $\l |\r''$,
we can express eq.\ebtwentythree\ as,
$$
\l\tPh^c_{2,\bt}| b_0^- b_0^+ (\tL_0^+)^{-1}|\tPh^c_{2,\at}\r''
-\l\tPh^c_{2,\bt}| b_0^- b_0^+ (\tL_0^+)^{-1}|\tPh^c_{2,\at}\r
+\co(\lambda^2)
\eqn\ebtwentyfour
$$
Thus the terms inside the curly bracket in eq.\ebtwentytwo\ take the form:
$$\eqalign{
&\l\tPh^c_{2,\bt}| b_0^- b_0^+ ((\tL_0^+)^{-1} -
(L_0^+)^{-1})|\tPh^c_{2,\at}\r'' \cr
& - \l\tPh^c_{2,\bt}| b_0^- b_0^+ ((\tL_0^+)^{-1} -
(L_0^+)^{-1})|\tPh^c_{2,\at}\r + \co(\lambda^2)\cr
}
\eqn\ebtwentyfive
$$
Note that $(\tL_0^+)^{-1}$ and $(L_0^+)^{-1}$ differ by a term of order
$\lambda$.
Also, $\l |\r''$ and $\l |\r$ differ by a term of order $\lambda$.
Thus the expression given in eq.\ebtwentyfive\ is of order $\lambda^2$.

This shows that if $\tA^\thr_{\mt\nt\kt}$ and
$\tA^\thr_{\st\tt\kt}$ vanish, then the left hand side of eq.\efoursix\
vanishes.
This, in turn, implies that the left hand side of eq.\efoursix\ must be of
the form given in eq.\efourseven.
Although the quantities $\Sigma^\one_{\kt\mt\nt}$ and hence
$S^\one_{\kt\mt\nt}$ may be determined by careful analysis (similar to the
one in ref.\BACKTWO\ which determined $\l\tPh^c_{2,\kt}|
K|\tPh_{2,l_2}\r$), we shall not carry out that analysis here.

\Appendix{C}

In this appendix we shall derive an expression for $T^\one_{\bt\kt}$
appearing on the right hand side of eq.\efivefifteen.
{}From eqs.\efivetwo, \efivethree, we get,
$$
T^\one_{\bt\kt} = \lim_{\lambda\to 0}(K^\zero_{\bt\kt} + W^\TWO_{\bt\kt})
\eqn\efiveeighteen
$$
where $K^\zero$ is defined in eqs.\ebone, \ebthree, and $W^{[n]}$ is
defined through the relation,
$$
V^{[n]}_{sr} = \delta_{sr}+\lambda W^{[n]}_{sr} +\co(\lambda^2)
\eqn\efiveseventeen
$$
We now use eqs.\ebfour, \ebtwenty, and the relation,
$$
b_0^- |\tPh^c_{2,\bt}\r = - \tqb b_0^- |\tPh^c_{3,\bt}\r
\eqn\efivetwenty
$$
to express $K^\zero_{\bt\kt}$ as,
$$
K^\zero_{\bt\kt} = \l\tPh^c_{3,\bt}|[\tqb, K]|\tPh_{2,\kt}\r''
={1\over\lambda} \l\tPh^c_{3,\bt}|(\tqb -\hqb)|\tPh_{2,\kt}\r''
\eqn\efivetwentyone
$$

Proof of eq.\efivetwenty: To prove this equation let us note that
$|\tilde\Phi^c_{2,\bt}\r$ is defined through the equations:
$$
\l\tPh^c_{2,\bt}|\tPh_{2,\ao}\r'' =0,~~ \l\tPh^c_{2,\bt}|\tPh_{2,\kt}\r''
=0, ~~ \l\tPh^c_{2,\bt}|\tPh_{2,\at}\r =\delta_{\at\bt}
\eqn\efivetwentya
$$
It is clear that $|\tPh^c_{2,\bt}\r$ given in eq.\efivetwenty\ satisfy the
first two sets of equations given in eq.\efivetwentya.
To verify the last set of equations in eq.\efivetwentya\ let us note that
with $|\tPh^c_{2,\bt}\r$ as defined in eq.\efivetwenty, we get,
$$\eqalign{
\l\tPh^c_{2,\bt}|\tPh_{2,\at}\r'' =& -\l\tqb b_0^-\tPh^c_{3,\bt} | c_0^-
|\tPh_{2,\at}\r'' \cr
=&\l\tPh^c_{3,\bt}|\tqb b_0^- c_0^- |\tPh_{2,\at}\r''
=\l\tPh^c_{3,\bt}|\tPh_{3,\at}\r'' =\delta_{\at\bt} \cr
}
\eqn\efivetwentyb
$$
This shows that $|\tPh^c_{2,\bt}\r$ defined in eq.\efivetwenty\  also
satisfies the last set of equations \efivetwentya.

Let us now compute $W^\TWO_{\bt\kt}$.
{}From eqs.\etwosevena\ and \efiveseventeen\ we see that,
$$
|\tPh_{n,r}\r = |\Phi_{n,r}\r +\lambda W^{[n]}_{sr}|\Phi_{n,s}\r
+\co(\lambda^2)
\eqn\efivetwentytwo
$$
Let us divide the basis $\{|\Phi_{n,r}\r\}$ into physical $\{|\Phi_{n,
k_n}\r\}$, unphysical
$\{|\Phi_{n,\alpha_n}\r\}$, and pure gauge $\{|\Phi_{n,\alpha_{n-1}}\r
= Q_B |\Phi_{n-1,\alpha_{n-1}}\r\}$ as in eqs.\ethreeone-\ethreethree,
with respect to the BRST charrge $Q_B$, and let $\{\l \Phi^c_{n,r}|\}$ be
the conjugate basis defined with respect to the BPZ inner product $\l |
\r$ in CFT.
In this case, $\{b_0^-|\Phi^c_{n, \alpha_n}\r\}$, $\{b_0^-|\Phi^c_{n,
k_n}\r\}$ and $\{b_0^-|\Phi^c_{n, \alpha_{n-1}}\r\}$ correspond to the
basis of pure gauge, physical, and unphysical states respectively of ghost
number $(5-n)$\BACKTWO.

We now note that in the generic case, the basis of unphysical states can
be taken to be the same in CFT and CFT$''$; these are the states that are
not annihilated by $Q_B$ or $\tqb$.
Using this, let us take,
$$
b_0^- |\Phi^c_{n, \alpha_{n-1}}\r = b_0^- |\tPh^c_{n, \alpha_{n-1}}\r
\eqn\efivetwentythree
$$
Using eq.\efivetwentythree\ and the analog of eq.\efivetwenty\ in CFT, we
get,
$$
b_0^-|\Phi^c_{2,\at}\r = - Q_B b_0^- |\Phi^c_{3,\at}\r = - Q_B b_0^-
|\tPh^c_{3,\at}\r
\eqn\efivetwentyfour
$$
Eq.\efivetwentytwo\ now gives,
$$
\lambda W^\TWO_{\bt\kt} =\l\Phi^c_{2,\bt}|\tPh_{2,\kt}\r+\co(\lambda^2)
=\l\tPh^c_{3,\bt}|
Q_B|\tPh_{2,\kt}\r +\co(\lambda^2)
\eqn\efivetwentyfive
$$
Using eqs.\efiveeighteen, \efivetwentyone, and \efivetwentyfive, and the
relation $\tqb |\tPh_{2,\kt}\r =0$, we get,
$$
T^\one_{\bt\kt} =\lim_{\lambda\to 0}\Big( {1\over\lambda}
\l\tPh^c_{3,\bt}|(Q_B -\hqb)
|\tilde\Phi_{2,\kt}\r \Big)
\eqn\efivetwentysix
$$
Using eqs.\etwonineteen\ and \etwotwentysix\ and that $|\tPh_{2,\kt}\r=
c_1\bc_1 |\vp\r+\co(\lambda)$, we get,
$$
T^\one_{\bt\kt} = -\sqrt 2 \{(\tPh^c_{3,\bt})(c_0^-c_1\bc_1|\vp\r)
(c_0^-c_1\bc_1|\vp\r)  \}''
\eqn\efivetwentyseven
$$

\refout

\end